\documentclass[11pt,a4paper]{scrartcl}
\pdfoutput=1
\usepackage{jheppub}
\usepackage[utf8]{inputenc}
\usepackage[T1]{fontenc}
\usepackage[british]{babel}
\usepackage{amsmath,amssymb,amsfonts}
\usepackage[usenamnes,dvipsnames]{xcolor}
\usepackage{epsfig}
\usepackage{epstopdf}
\usepackage{graphicx,psfrag}
\usepackage{tikz}
\usepackage{tabu}
\usepackage[title]{appendix}

\usetikzlibrary{arrows,decorations.pathmorphing,backgrounds}
\usetikzlibrary{decorations.pathreplacing,shapes}
\usetikzlibrary{positioning,calc,fit,petri,shadows,matrix}

\tikzset{
  baseline={([yshift=-.5ex]current bounding box.center)},
  skip loop/.style={to path={-- ++(0,#1) -| (\tikztotarget)}}
}

\def\clap#1{\hbox to 0pt{\hss#1\hss}}

\def\mathrlap{\mathpalette\mathrlapinternal}
\def\mathclap{\mathpalette\mathclapinternal}

\def\mathrlapinternal#1#2{%
	\rlap{$\mathsurround=0pt#1{#2}$}}

\def\mathclapinternal#1#2{%
	\clap{$\mathsurround=0pt#1{#2}$}}

\newenvironment{midtikzpicture}
{\begin{tikzpicture}[baseline={([yshift=-.5ex]current bounding box.center)}]}
{\end{tikzpicture}}

\newcommand{\midtikz}{\tikz[baseline={([yshift=-.5ex]current bounding box.center)}]}

\usetikzlibrary{positioning,arrows}
\usetikzlibrary{decorations.pathmorphing}
\usetikzlibrary{decorations.markings}
\usetikzlibrary{backgrounds}
\usetikzlibrary{shapes}
\usetikzlibrary{positioning,calc,fit,petri,shadows,matrix}

\tikzset{
  skip loop/.style={to path={-- ++(0,#1) -| (\tikztotarget)}}
}

\tikzset{
    position/.style args={#1:#2 from #3}{
        at=(#3.#1), anchor=#1+180, shift=(#1:#2)
    }
}

\tikzset{
  circ/.style={%
    black,thick,circle,draw,inner sep=2pt,scale=0.75
  },
  dot/.style={
    black,fill,circle,inner sep=0pt,minimum size=6pt
  },
  smalldot/.style={
    black,fill,circle,inner sep=0pt,minimum size=3.5pt
  },
  inner/.style={
    black,fill,circle,inner sep=0pt,minimum size=3pt
  },
  outer/.style={
    black,draw,thick,circle,inner sep=0pt,minimum size=6pt
  },
  directed/.style={postaction={decorate},
    decoration={markings,mark=at position .65 with{\arrow{stealth}}}},
}

\newcommand{\beq}{\begin{equation}}
\newcommand{\eeq}{\end{equation}}
\newcommand{\bea}{\begin{eqnarray}}
\newcommand{\eea}{\end{eqnarray}}
\newcommand{\nn}{\nonumber}
\newcommand{\eq}{Eq.~}

\newcommand{\fig}{Fig.~}

\DeclareMathOperator{\tr}{tr}

\def\lsi{\raise0.3ex\hbox{$<$\kern-0.75em\raise-1.1ex\hbox{$\sim$}}}
\def\gsi{\raise0.3ex\hbox{$>$\kern-0.75em\raise-1.1ex\hbox{$\sim$}}}
\newcommand{\losim}{\mathop{\lsi}}

\title{Equation of state for cold and dense heavy QCD}
\author{Jonas Glesaaen, Mathias Neuman, Owe Philipsen}
\affiliation{Institut f\"ur Theoretische Physik,
Goethe-Universit\"at Frankfurt, \\ Max-von-Laue-Str. 1, 60438 Frankfurt am Main, Germany}

\emailAdd{glesaaen, philipsen@th.physik.uni-frankfurt.de}

\abstract{
A previously derived three-dimensional effective lattice theory describing the thermodynamics 
of QCD with heavy quarks in the cold and dense region is extended through order $\sim u^5\kappa^8$ in 
the combined character and hopping expansion of the original four-dimensional Wilson action.  
The systematics of the effective theory is investigated to determine its range of validity in parameter 
space. We demonstrate the severe cut-off effects due to lattice saturation, which afflict any lattice results at finite
baryon density independent of the sign problem or the quality of effective theories, 
and which have to be removed by continuum extrapolation. 
We then show how the effective theory can be solved analytically by means of a linked cluster expansion, 
which is completely unaffected by the sign problem,
in quantitative agreement with numerical simulations. As an application, we compute the cold 
nuclear equation of state of heavy QCD. Our continuum extrapolated result is consistent with a polytropic equation
of state for non-relativistic fermions. 
}

\begin{document}
\maketitle


\newpage
\section{Introduction}

A fully non-perturbative description of QCD at finite baryon density, and in particular of nuclear
matter at zero temperature, remains an outstanding challenge because of the infamous
``sign problem'' of lattice QCD, which prohibits direct Monte Carlo simulations of that regime.
Approximate methods like reweighting, Taylor expansion about quark chemical potential $\mu=0$
or analytic continuation of results at imaginary chemical potential, for which there is
no sign problem, all require $\mu/T\lsi 1$ in order to work \cite{sign}.
So far, no signal of a critical point or a first order phase transition has been found in this 
controlled region, where the different approaches are in quantitative agreement.  
Simulations by complex Langevin algorithms do not suffer 
from the sign problem and a lot of progress in extending their range of operation 
has been achieved \cite{langevin1,langevin2}. Nevertheless, coarser lattices still appear inaccessible
and so far mostly the heavy dense limit of QCD has been studied \cite{lang_hop}. 
Up to now, no non-analytic phase transitions have been reported from this approach.

It is therefore desirable to further develop alternative approaches where the sign problem is 
fully controlled, even if those are restricted to certain parameter regions of QCD. 
Chiral fermions reformulated to a flux representation can be simulated in the 
strong coupling regime by means of 
a worm algorithm, and gauge corrections can be successively included \cite{wolfgang}. Also 
analytic expansion methods are attempted in that regime \cite{tomboulis}.
Here we pursue the complementary approach and consider QCD thermodynamics 
with heavy quarks but much closer to the continuum. This situation is described by a 
3d effective theory derived by a combined character and hopping expansion and features
a mild sign problem only, allowing for the simulation of real chemical potentials and the mapping of
both the hot and cold regions of the phase diagram \cite{kappa,silver,densek4}. This approach 
has also been successfully tested against two-colour-QCD \cite{smekal}, where there is no sign problem.

\begin{figure}
\centerline{
\includegraphics[width=0.5\textwidth]{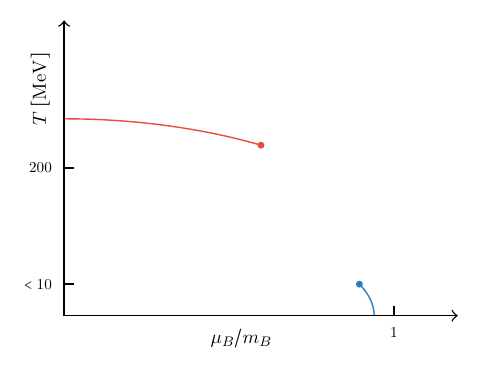}
}
\caption[]{The phase diagram of QCD with very heavy quarks.}
\label{fig:zn}
\end{figure}

The phase diagram of QCD with heavy quarks is sketched in \fig\ref{fig:zn}. At zero density
there is a first order deconfinement transition related to the spontaneous breaking of the 
centre symmetry in pure gauge theory. Real chemical potential for quarks weakens that transition,
which then features a critical end point. The thermal phase transition and the dependence of 
the critical end point on the quark or pion mass has been calculated in \cite{kappa} on the lattice.
The same picture emerges in studies of continuum Polyakov loop models \cite{sasaki} or Dyson-Schwinger
equations with heavy quarks \cite{fischer}.
At low temperatures and higher chemical potentials, also 
the nuclear liquid gas transition is accessible \cite{silver}.
The temperature of its critical endpoint is of the order of the nuclear binding energy and 
also depends on the quark mass. 
In the infinite mass limit it moves to zero temperature \cite{densek4}.

In the present work, we extend the results of \cite{densek4} in two ways.
First, we push the derivation of the effective action for the cold and dense regime through order
$u^5\kappa^8$ to leading order in $N_{\tau}^{-1}$. Second and most importantly, we apply linked cluster expansion
methods \cite{wortis} to our effective theory and demonstrate that its thermodynamic functions
and equation of state can be computed entirely analytically in the domain of its validity. We then devise
a resummation scheme to sum up a particular class of diagrams and finally compute the equation of
state for heavy nuclear matter.

\section{The effective theory}

\subsection{Derivation}

The derivation of the effective theory has been discussed in previous publications 
\cite{ym, kappa, densek4},
so we only outline the
procedure and give our results.
Starting point is lattice QCD with the Wilson plaquette and fermion actions on
an $N_s^3\times N_\tau$ lattice,
\begin{eqnarray}
Z=\int[dU_\mu]\;\exp\left[-S_g\right]\prod_{f=1}^{N_f}\det\left[Q^f\right]\;,\qquad
-S_g=\frac{\beta}{2N_c}\sum_p\left[\tr\,   U_p+\tr\,   U_p^\dagger\right]\;,
\end{eqnarray}
with elementary plaquettes $U_p$, the quark hopping matrix for the flavour $f$,
\begin{eqnarray}
&&(Q^f)^{ab}_{\alpha\beta,xy}=\delta^{ab}\delta_{\alpha\beta}\delta_{xy}\\ \hspace*{-1.5cm}
&&-\kappa_f\sum_{\nu=0}^3\left[e^{a\mu_f\delta_{\nu0}}(1+\gamma_\nu)_{\alpha\beta}U_\nu^{ab}(x)
\delta_{x,y-\hat{\nu}}+e^{-a\mu_f\delta_{\nu0}}(1-\gamma_\nu)_{\alpha\beta}U_{-\nu}^{ab}(x)
\delta_{x,y+\hat{\nu}}\right]\;,
\;\nonumber
\end{eqnarray}
and $U_{-\nu}^{ab}(x) = U_{\nu}^{\dagger ab}(x-\hat{\nu})$. We denote colour indices with Latin characters and Dirac indices with
Greek characters throughout the article.  The effective action is defined by integrating over the spatial link variables,
\begin{align}
  Z &=\int[dU_0]\;\exp[-S_\mathrm{eff}]\;,\\
  \exp[-S_{\mathrm{eff}}] &\equiv \int[dU_k]\exp\left[-S_g\right]\prod_{f=1}^{N_f}\det\left[Q^f\right]\;, \label{eq:eff_action_def}\\
  S_{\mathrm{eff}} &= \sum_{i=0}^{\infty}S^g_{i}(\beta, \kappa_f,N_{\tau};W) + \sum_{i=0}^{\infty} S^f_{i}(\beta, N_{\tau}, \kappa_f, \mu_f;W) \;, 
\label{eq_defeffth}
\end{align}
and we have split the effective action 
into contributions coming from the pure gauge theory and the fermion
determinant. We now specify $N_f=2$ degenerate quarks with $\kappa_u=\kappa_d=\kappa$.
In our approach we approximate the exponential in \eqref{eq:eff_action_def} by truncated
expansions in the fundamental character of the gauge group, $u(\beta)=\beta/18+O(\beta^2)$ \cite{Montvay:1994cy},
and the hopping parameter $\kappa = (2 a m + 8)^{-1}$. After the expansion, the gauge integration can be done 
analytically. The resulting effective theory is three-dimensional and only depends on 
temporal Wilson lines or, equivalently, Polyakov loops,
\begin{equation}
  W(\vec{x})=\prod_{\tau=1}^{N_\tau} U_0({\vec{ x}},\tau),\quad L(\vec{x})=\tr W(\vec{x})\;,
\end{equation}
with $n$-point interaction terms containing all powers of fields at all separations.
Note that the static determinant can be computed exactly, i.e.~hops in the temporal
directions are included to all orders,
\beq
\det[Q_{\mathrm{stat}}]= \prod_{\vec{x}} 
\det \Big[1+h_1W({\vec{x}})\Big]^2
\det \Big[1+\bar{h}_1W^{\dagger}({\vec{x}})\Big]^2\;,
\label{q_static}
\eeq 
with the one-point coupling constants to leading order
\beq
h_1=(2 \kappa e^{a \mu})^{N_{\tau}},\quad \bar{h}_1=(2 \kappa e^{-a \mu})^{N_{\tau}}\;.
\eeq
The static determinant has a particle-hole symmetry 
about half-filling akin to the Hubbard model \cite{rindli}.
The expansion is then in spatial hops of the remaining
kinetic determinant,
\begin{eqnarray}
\det[Q]&\equiv&\det[Q_{\mathrm{stat}}]\det[Q_{\mathrm{kin}}]\;.
\label{eq_detqkin}
\end{eqnarray}
For our physics region of interest, the cold and dense regime $\mu\gg T$, 
considerable simplifications arise.
At fixed lattice spacing the zero temperature limit corresponds to $N_\tau\rightarrow \infty$.
The centre-symmetric couplings, $\lambda_i$, have been calculated in previous publications  
and tested against the full Yang-Mills theory \cite{ym,Bergner:2013qaa,Bergner:2015rza}.
They are suppressed as 
$\lambda_i\sim u^{nN_\tau}$ with $n\geq 1$ and $u(\beta)<1$ always.
In this work we employ $\beta\leq 6.2$ and $N_\tau\geq 116$, such that $\lambda_1 \losim 10^{-18}$
and $\lambda_i \leq \lambda_1$. Thus the pure gauge sector plays no role in the cold and dense
regime and can be safely neglected. Similarly, $\bar{h}_1\rightarrow 0$ in the zero temperature limit.
The summation 
of all temporal windings produces the basic building blocks of the effective action,
\begin{equation}
  W_{n,m}(\vec{x}) =\tr \, \frac{\big(h_1 W(\vec{x})\big)^m}{\big(1+h_1 W(\vec{x})\big)^n} \, .
\end{equation}
We have calculated the effective action through order $\kappa^8 u^5$ in the low temperature limit, 
i.e.~to the leading power of
$N_\tau$. Because of its length we will give the result here in a compact, graphical representation 
and relegate the full expression to appendix \ref{app:action}.  We symbolise
factors of $W_{n,m}(\vec{x})$ by vertices, where $n$ is the number of bonds entering a vertex, and $m$ is the number indicated on
the node. Furthermore, vertices which are connected by one or more bonds are nearest neighbours on the lattice.
{\allowdisplaybreaks%
\begin{align} \label{eq:effective_action}
  S_{\mathrm{eff}} &=
  h_2 N_f \sum_{\mathrm{dof}} \; \begin{midtikzpicture}
    \node[circ] (n1) {1};
    \node[circ] (n2) at ([shift={(270:.75)}] n1) {1}
      edge[thick] (n1);
  \end{midtikzpicture}
  \; - h_2^2 N_f\sum_{\mathrm{dof}}  \: \begin{midtikzpicture}
    \node[circ] (n1) {1};
    \node[circ] (n2) at ([shift={(60:{sqrt(2/3)})}] n1) {1}
      edge[thick] (n1);
    \node[circ] (n3) at ([shift={(300:{sqrt(2/3)})}] n2) {1}
      edge[thick] (n2);
  \end{midtikzpicture}
  - h_2^2 N_f^2\sum_{\mathrm{dof}} \: \begin{midtikzpicture}
    \node[circ] (n1) {1};
    \node[circ] (n2) at ([shift={(270:.75)}] n1) {1}
      edge[thick,bend left=45] (n1)
      edge[thick,bend right=45] (n1);
  \end{midtikzpicture}
  + h_2^3 N_f\sum_{\mathrm{dof}} \: \begin{midtikzpicture}
    \node[circ] (n1) {1};
    \node[circ] (n2) at ([shift={(60:{sqrt(2/3)})}] n1) {1}
      edge[thick] (n1);
    \node[circ] (n3) at ([shift={(300:{sqrt(2/3)})}] n2) {1}
      edge[thick] (n2);
    \node[circ] (n4) at ([shift={(60:{sqrt(2/3)})}] n3) {1}
      edge[thick] (n3);
  \end{midtikzpicture} \nonumber\\
  &+\frac{1}{3} h_2^3 N_f \sum_{\mathrm{dof}} \Bigg(\; \begin{midtikzpicture}
    \node[circ] (n1) {1};
    \node[circ] at ([shift={(-30:.5)}] n1) {1}
      edge[thick] (n1);
    \node[circ] at ([shift={(90:.5)}] n1) {1}
      edge[thick] (n1);
    \node[circ] at ([shift={(210:.5)}] n1) {1}
      edge[thick] (n1);
  \end{midtikzpicture} 
  \;-\;\begin{midtikzpicture}
    \node[circ] (n1) {2};
    \node[circ] at ([shift={(-30:.5)}] n1) {1}
      edge[thick] (n1);
    \node[circ] at ([shift={(90:.5)}] n1) {1}
      edge[thick] (n1);
    \node[circ] at ([shift={(210:.5)}] n1) {1}
      edge[thick] (n1);
  \end{midtikzpicture} \;\Bigg)
  +2 h_2^3 N_f^2 \sum_{\mathrm{dof}} \Bigg(\; \begin{midtikzpicture}
    \node[circ] (n1) {1};
    \node[circ] (n2) at ([shift={(60:.75)}] n1) {1}
      edge[thick,bend left=30] (n1)
      edge[thick,bend right=30] (n1);
    \node[circ] at ([shift={(300:.75)}] n2) {1}
      edge[thick] (n2);
  \end{midtikzpicture} 
  \;-\; \begin{midtikzpicture}
    \node[circ] (n1) {1};
    \node[circ] (n2) at ([shift={(60:.75)}] n1) {2}
      edge[thick,bend left=30] (n1)
      edge[thick,bend right=30] (n1);
    \node[circ] at ([shift={(300:.75)}] n2) {1}
      edge[thick] (n2);
  \end{midtikzpicture} \Bigg) \nonumber\\
  &+\frac{1}{6} h_2^3 N_f \sum_{\mathrm{dof}} \Bigg(\; \begin{midtikzpicture}
    \node[circ] (n1) {1};
    \node[circ] at ([shift={(270:.75)}] n1) {1}
      edge[thick,bend left=45] (n1)
      edge[thick,bend right=45] (n1)
      edge[thick] (n1);
  \end{midtikzpicture} 
  \;-\; \begin{midtikzpicture}
    \node[circ] (n1) {2};
    \node[circ] at ([shift={(270:.75)}] n1) {2}
      edge[thick,bend left=45] (n1)
      edge[thick,bend right=45] (n1)
      edge[thick] (n1);
  \end{midtikzpicture} \Bigg)
  - \frac{4}{3} h_2^3 N_f^3 \sum_{\mathrm{dof}} \; \begin{midtikzpicture}
    \node[circ] (n1) {1};
    \node[circ] at ([shift={(270:.75)}] n1) {2}
      edge[thick,bend left=45] (n1)
      edge[thick,bend right=45] (n1)
      edge[thick] (n1);
  \end{midtikzpicture} 
  - h_2^4 N_f\sum_{\mathrm{dof}} \: \begin{midtikzpicture}
    \node[circ] (n1) {1};
    \node[circ] (n2) at ([shift={(60:{sqrt(2/3)})}] n1) {1}
      edge[thick] (n1);
    \node[circ] (n3) at ([shift={(300:{sqrt(2/3)})}] n2) {1}
      edge[thick] (n2);
    \node[circ] (n4) at ([shift={(60:{sqrt(2/3)})}] n3) {1}
      edge[thick] (n3);
    \node[circ] (n5) at ([shift={(300:{sqrt(2/3)})}] n4) {1}
      edge[thick] (n4);
  \end{midtikzpicture} \nonumber\\
  &- \frac{1}{12} h_2^4 N_f\sum_{\mathrm{dof}} \Bigg(\; \begin{midtikzpicture}
    \node[circ] (n1) {1};
    \node[circ] at ([shift={(45:.5)}] n1) {1}
      edge[thick] (n1);
    \node[circ] at ([shift={(135:.5)}] n1) {1}
      edge[thick] (n1);
    \node[circ] at ([shift={(225:.5)}] n1) {1}
      edge[thick] (n1);
    \node[circ] at ([shift={(315:.5)}] n1) {1}
      edge[thick] (n1);
  \end{midtikzpicture}
  \;-2\; \begin{midtikzpicture}
    \node[circ] (n1) {2};
    \node[circ] at ([shift={(45:.5)}] n1) {1}
      edge[thick] (n1);
    \node[circ] at ([shift={(135:.5)}] n1) {1}
      edge[thick] (n1);
    \node[circ] at ([shift={(225:.5)}] n1) {1}
      edge[thick] (n1);
    \node[circ] at ([shift={(315:.5)}] n1) {1}
      edge[thick] (n1);
  \end{midtikzpicture}
  \;+\; \begin{midtikzpicture}
    \node[circ] (n1) {3};
    \node[circ] at ([shift={(45:.5)}] n1) {1}
      edge[thick] (n1);
    \node[circ] at ([shift={(135:.5)}] n1) {1}
      edge[thick] (n1);
    \node[circ] at ([shift={(225:.5)}] n1) {1}
      edge[thick] (n1);
    \node[circ] at ([shift={(315:.5)}] n1) {1}
      edge[thick] (n1);
  \end{midtikzpicture} \Bigg)
  -  h_2^4 N_f\sum_{\mathrm{dof}} \Bigg(\; \begin{midtikzpicture}
    \node[circ] (n1) {1};
    \node[circ] (n2) at ([shift={(60:{sqrt(1/3)})}] n1) {1}
      edge[thick] (n1);
    \node[circ] (n3) at ([shift={(300:{sqrt(1/3)})}] n2) {1}
      edge[thick] (n2);
    \node[circ] (n4) at ([shift={(60:{sqrt(1/3)})}] n3) {1}
      edge[thick] (n3);
    \node[circ] (n5) at ([shift={(0:.5)}] n3) {1}
      edge[thick] (n3);
  \end{midtikzpicture}
  \;-\; \begin{midtikzpicture}
    \node[circ] (n1) {1};
    \node[circ] (n2) at ([shift={(60:{sqrt(1/3)})}] n1) {1}
      edge[thick] (n1);
    \node[circ] (n3) at ([shift={(300:{sqrt(1/3)})}] n2) {2}
      edge[thick] (n2);
    \node[circ] (n4) at ([shift={(60:{sqrt(1/3)})}] n3) {1}
      edge[thick] (n3);
    \node[circ] (n5) at ([shift={(0:.5)}] n3) {1}
      edge[thick] (n3);
  \end{midtikzpicture} \Bigg) \nonumber\\
  &- h_2^4 N_f^2 \sum_{\mathrm{dof}} \Bigg(\; \begin{midtikzpicture}
    \node[circ] (n1) {1};
    \node[circ] at ([shift={(-30:.5)}] n1) {1}
      edge[thick] (n1);
    \node[circ] at ([shift={(90:.65)}] n1) {1}
      edge[thick,bend left=30] (n1)
      edge[thick,bend right=30] (n1);
    \node[circ] at ([shift={(210:.5)}] n1) {1}
      edge[thick] (n1);
  \end{midtikzpicture} 
  -4\begin{midtikzpicture}
    \node[circ] (n1) {2};
    \node[circ] at ([shift={(-30:.5)}] n1) {1}
      edge[thick] (n1);
    \node[circ] at ([shift={(90:.65)}] n1) {1}
      edge[thick,bend left=30] (n1)
      edge[thick,bend right=30] (n1);
    \node[circ] at ([shift={(210:.5)}] n1) {1}
      edge[thick] (n1);
  \end{midtikzpicture}
  +\begin{midtikzpicture}
    \node[circ] (n1) {3};
    \node[circ] at ([shift={(-30:.5)}] n1) {1}
      edge[thick] (n1);
    \node[circ] at ([shift={(90:.65)}] n1) {1}
      edge[thick,bend left=30] (n1)
      edge[thick,bend right=30] (n1);
    \node[circ] at ([shift={(210:.5)}] n1) {1}
      edge[thick] (n1);
  \end{midtikzpicture} \;\Bigg)
  - h_2^4 N_f^2 \sum_{\mathrm{dof}} \begin{midtikzpicture}
    \node[circ] (n1) {1};
    \node[circ] (n2) at (0.65,0) {1}
      edge[thick] (n1);
    \node[circ] (n3) at (0.65,0.65) {1}
      edge[thick] (n2);
    \node[circ] (n4) at (0,0.65) {1}
      edge[thick] (n3)
      edge[thick] (n1);
  \end{midtikzpicture} \nonumber\\
  &-2 h_2^4 N_f^2 \sum_{\mathrm{dof}} \Bigg( \; \begin{midtikzpicture}
    \node[circ] (n1) {1};
    \node[circ] (n2) at ([shift={(60:.75)}] n1) {1}
      edge[thick,bend left=30] (n1)
      edge[thick,bend right=30] (n1);
    \node[circ] (n3) at ([shift={(300:.75)}] n2) {1}
      edge[thick] (n2);
    \node[circ] at ([shift={(60:.75)}] n3) {1}
      edge[thick] (n3);
  \end{midtikzpicture}
  \;-\; \begin{midtikzpicture}
    \node[circ] (n1) {1};
    \node[circ] (n2) at ([shift={(60:.75)}] n1) {2}
      edge[thick,bend left=30] (n1)
      edge[thick,bend right=30] (n1);
    \node[circ] (n3) at ([shift={(300:.75)}] n2) {1}
      edge[thick] (n2);
    \node[circ] at ([shift={(60:.75)}] n3) {1}
      edge[thick] (n3);
  \end{midtikzpicture} \Bigg)
  - h_2^4 N_f^2 \sum_{\mathrm{dof}} \Bigg( \; \begin{midtikzpicture}
    \node[circ] (n1) {1};
    \node[circ] (n2) at ([shift={(65:{sqrt(2/3)})}] n1) {1}
      edge[thick] (n1);
    \node[circ] (n3) at ([shift={(-65:{sqrt(2/3)})}] n2) {1}
      edge[thick,bend left=30] (n2)
      edge[thick,bend right=30] (n2);
    \node[circ] (n4) at ([shift={(65:{sqrt(2/3)})}] n3) {1}
      edge[thick] (n3);
  \end{midtikzpicture}
  \;-2\; \begin{midtikzpicture}
    \node[circ] (n1) {1};
    \node[circ] (n2) at ([shift={(65:{sqrt(2/3)})}] n1) {1}
      edge[thick] (n1);
    \node[circ] (n3) at ([shift={(-65:{sqrt(2/3)})}] n2) {2}
      edge[thick,bend left=30] (n2)
      edge[thick,bend right=30] (n2);
    \node[circ] (n4) at ([shift={(65:{sqrt(2/3)})}] n3) {1}
      edge[thick] (n3);
  \end{midtikzpicture}
  \;+\; \begin{midtikzpicture}
    \node[circ] (n1) {1};
    \node[circ] (n2) at ([shift={(65:{sqrt(2/3)})}] n1) {2}
      edge[thick] (n1);
    \node[circ] (n3) at ([shift={(-65:{sqrt(2/3)})}] n2) {2}
      edge[thick,bend left=30] (n2)
      edge[thick,bend right=30] (n2);
    \node[circ] (n4) at ([shift={(65:{sqrt(2/3)})}] n3) {1}
      edge[thick] (n3);
  \end{midtikzpicture} \Bigg) \nonumber\\
  &-\frac{1}{3} h_2^4 N_f \sum_{\mathrm{dof}} \Bigg( \; \begin{midtikzpicture}
    \node[circ] (n1) {1};
    \node[circ] (n2) at ([shift={(60:.75)}] n1) {1}
      edge[thick,bend left=30] (n1)
      edge[thick,bend right=30] (n1)
      edge[thick] (n1);
    \node[circ] (n3) at ([shift={(300:.75)}] n2) {1}
      edge[thick] (n2);
  \end{midtikzpicture}
  \;-2\; \begin{midtikzpicture}
    \node[circ] (n1) {1};
    \node[circ] (n2) at ([shift={(60:.75)}] n1) {2}
      edge[thick,bend left=30] (n1)
      edge[thick,bend right=30] (n1)
      edge[thick] (n1);
    \node[circ] (n3) at ([shift={(300:.75)}] n2) {1}
      edge[thick] (n2);
  \end{midtikzpicture}
  \;+2\; \begin{midtikzpicture}
    \node[circ] (n1) {2};
    \node[circ] (n2) at ([shift={(60:.75)}] n1) {2}
      edge[thick,bend left=30] (n1)
      edge[thick,bend right=30] (n1)
      edge[thick] (n1);
    \node[circ] (n3) at ([shift={(300:.75)}] n2) {1}
      edge[thick] (n2);
  \end{midtikzpicture}
  \;-\; \begin{midtikzpicture}
    \node[circ] (n1) {2};
    \node[circ] (n2) at ([shift={(60:.75)}] n1) {3}
      edge[thick,bend left=30] (n1)
      edge[thick,bend right=30] (n1)
      edge[thick] (n1);
    \node[circ] (n3) at ([shift={(300:.75)}] n2) {1}
      edge[thick] (n2);
  \end{midtikzpicture} \Bigg) \nonumber\\
  &+\frac{4}{3} h_2^4 N_f^3 \sum_{\mathrm{dof}} \Bigg( \; \begin{midtikzpicture}
    \node[circ] (n1) {2};
    \node[circ] (n2) at ([shift={(60:.75)}] n1) {1}
      edge[thick,bend left=30] (n1)
      edge[thick,bend right=30] (n1)
      edge[thick] (n1);
    \node[circ] (n3) at ([shift={(300:.75)}] n2) {1}
      edge[thick] (n2);
  \end{midtikzpicture}
  \;-2\; \begin{midtikzpicture}
    \node[circ] (n1) {2};
    \node[circ] (n2) at ([shift={(60:.75)}] n1) {2}
      edge[thick,bend left=30] (n1)
      edge[thick,bend right=30] (n1)
      edge[thick] (n1);
    \node[circ] (n3) at ([shift={(300:.75)}] n2) {1}
      edge[thick] (n2);
  \end{midtikzpicture}
  \;+2\; \begin{midtikzpicture}
    \node[circ] (n1) {1};
    \node[circ] (n2) at ([shift={(60:.75)}] n1) {2}
      edge[thick,bend left=30] (n1)
      edge[thick,bend right=30] (n1)
      edge[thick] (n1);
    \node[circ] (n3) at ([shift={(300:.75)}] n2) {1}
      edge[thick] (n2);
  \end{midtikzpicture}
  \;-\; \begin{midtikzpicture}
    \node[circ] (n1) {1};
    \node[circ] (n2) at ([shift={(60:.75)}] n1) {3}
      edge[thick,bend left=30] (n1)
      edge[thick,bend right=30] (n1)
      edge[thick] (n1);
    \node[circ] (n3) at ([shift={(300:.75)}] n2) {1}
      edge[thick] (n2);
  \end{midtikzpicture} \Bigg) \nonumber\\
  &-\bigg(\frac{1}{12} N_f + \frac{2}{3} N_f^3 \bigg) h_2^4 \sum_{\mathrm{dof}} \Bigg( \; \begin{midtikzpicture}
    \node[circ] (n1) {1};
    \node[circ] (n2) at ([shift={(60:.75)}] n1) {1}
      edge[thick,bend left=30] (n1)
      edge[thick,bend right=30] (n1);
    \node[circ] (n3) at ([shift={(300:.75)}] n2) {1}
      edge[thick,bend left=30] (n2)
      edge[thick,bend right=30] (n2);
  \end{midtikzpicture}
  \;-4\; \begin{midtikzpicture}
    \node[circ] (n1) {1};
    \node[circ] (n2) at ([shift={(60:.75)}] n1) {2}
      edge[thick,bend left=30] (n1)
      edge[thick,bend right=30] (n1);
    \node[circ] (n3) at ([shift={(300:.75)}] n2) {1}
      edge[thick,bend left=30] (n2)
      edge[thick,bend right=30] (n2);
  \end{midtikzpicture}
  \;+\; \begin{midtikzpicture}
    \node[circ] (n1) {1};
    \node[circ] (n2) at ([shift={(60:.75)}] n1) {3}
      edge[thick,bend left=30] (n1)
      edge[thick,bend right=30] (n1);
    \node[circ] (n3) at ([shift={(300:.75)}] n2) {1}
      edge[thick,bend left=30] (n2)
      edge[thick,bend right=30] (n2);
  \end{midtikzpicture} \Bigg)
  -\frac{2}{3} h_2^4 N_f^4 \sum_{\mathrm{dof}} \Bigg( \; \begin{midtikzpicture}
    \node[circ] (n1) {1};
    \node[circ] (n2) at (0,-.75) {3}
      edge[thick, bend right=45] (n1)
      edge[thick, bend left=45] (n1)
      edge[thick, bend right=15] (n1)
      edge[thick, bend left=15] (n1);
  \end{midtikzpicture}
  \;+2\; \begin{midtikzpicture}
    \node[circ] (n1) {2};
    \node[circ] (n2) at (0,-.75) {2}
      edge[thick, bend right=45] (n1)
      edge[thick, bend left=45] (n1)
      edge[thick, bend right=15] (n1)
      edge[thick, bend left=15] (n1);
  \end{midtikzpicture} \Bigg) \nonumber \\
  &-\frac{1}{12} h_2^4 N_f^2 \sum_{\mathrm{dof}} \Bigg( \; \begin{midtikzpicture}
    \node[circ] (n1) {1};
    \node[circ] (n2) at (0,-.75) {1}
      edge[thick, bend right=45] (n1)
      edge[thick, bend left=45] (n1)
      edge[thick, bend right=15] (n1)
      edge[thick, bend left=15] (n1);
  \end{midtikzpicture}
  \;+12\; \begin{midtikzpicture}
    \node[circ] (n1) {2};
    \node[circ] (n2) at (0,-.75) {2}
      edge[thick, bend right=45] (n1)
      edge[thick, bend left=45] (n1)
      edge[thick, bend right=15] (n1)
      edge[thick, bend left=15] (n1);
  \end{midtikzpicture}
  \;+\; \begin{midtikzpicture}
    \node[circ] (n1) {3};
    \node[circ] (n2) at (0,-.75) {3}
      edge[thick, bend right=45] (n1)
      edge[thick, bend left=45] (n1)
      edge[thick, bend right=15] (n1)
      edge[thick, bend left=15] (n1);
  \end{midtikzpicture} \Bigg)
  +\frac{2}{3} h_2^4 N_f^2 \sum_{\mathrm{dof}} \Bigg( \; \begin{midtikzpicture}
    \node[circ] (n1) {1};
    \node[circ] (n2) at (0,-.75) {2}
      edge[thick, bend right=45] (n1)
      edge[thick, bend left=45] (n1)
      edge[thick, bend right=15] (n1)
      edge[thick, bend left=15] (n1);
  \end{midtikzpicture}
  \;+\; \begin{midtikzpicture}
    \node[circ] (n1) {2};
    \node[circ] (n2) at (0,-.75) {3}
      edge[thick, bend right=45] (n1)
      edge[thick, bend left=45] (n1)
      edge[thick, bend right=15] (n1)
      edge[thick, bend left=15] (n1);
  \end{midtikzpicture} \Bigg) +\mathcal{O} \big(\kappa^{10}, \frac{1}{N_{\tau}}\big)
\end{align}
}%
The sums over the ``degrees of freedom'' constitute the traces in coordinate space.
The effective couplings to the order computed here are
\begin{align}
  h_1&= e^{N_{\tau} (a \mu + \log(2 \kappa))} \exp \Big[ 6 N_{\tau} \kappa^2 u \Big(\frac{\hskip5pt 1-u^{N_{\tau}-1}}{1-u} + 4u^4
  \nonumber \\
  &\hspace{5.5cm}-12\kappa^2+9\kappa^2 u +4\kappa^2 u^2-4\kappa^4\Big)\Big], \\
  h_2&= \frac{\kappa^2 N_{\tau}}{N_c} \Big[1 + 2\frac{\hskip4pt u - u^{N_{\tau}}}{1-u} + 8 u^5 + 16 \kappa^2 u^4 \Big]   \;.
\end{align}
Note that higher order corrections in $\kappa,u$ to $h_1$ and $h_2$ 
are subleading in the high $N_{\tau}$ limit.

\subsection{Observables}

We are interested in the thermodynamical functions, which are directly related to the partition 
function, in particular the baryon number density, pressure and energy density,
\bea
a^3n&=&\frac{1}{N_\tau N_s^3}\frac{\partial}{\partial a\mu}\ln Z\;,\\
a^4p&=&\frac{a^4T}{V}\ln Z=\frac{1}{N_\tau N_s^3}\ln Z\;,\\
a^4e&=&-\frac{a}{N_\tau N_s^3}\frac{\partial}{\partial a}\ln Z\bigg\vert_z\;.
\eea
Here $a$ is the lattice spacing and $z=\exp(a N_\tau\mu)$ the fugacity.

\subsection{Hadron masses and the physical scale}

In order to interpret the results in the following sections, we use the masses of 
mesons and baryons for $N_f=2$ to all orders in the hopping expansion \cite{smit} and to 
resummed next-to-leading order 
in the strong coupling expansion,
\bea
am_M&=& \text{ArcCosh}\Big[1+\frac{(M^2-4)(M^2-1)}{2M^2-3} \Big]-24\kappa^2\frac{u}{1-u}+\ldots\;,\label{eq:meson_mass}\\
am_B&=& \ln \Big[\frac{M^3(M^3-2)}{M^3-\frac{5}{4}} \Big]-18\kappa^2\frac{u}{1-u}+\ldots\;,
\label{eq:baryon_mass}
\eea
with $M=\frac{1}{2 \kappa}$. Explicit evaluation shows that these formulae are remarkably convergent
in the heavy mass regime for $\beta\losim6.2$, which we employ in our analysis.
Since heavy quarks have little influence on the running of the coupling we use the 
beta-function of pure gauge theory for the lattice spacing in units of the Sommer parameter, 
$a(\beta)/r_0$ with $r_0 = 0.5$ fm \cite{Necco:2001xg}. Temperature is then set via $T = (aN_\tau)^{-1}$.

\section{Simulation and systematics of the effective theory}

\subsection{Convergence of the effective action}

\begin{figure}[t]
\centerline{
\includegraphics[width=0.5\textwidth]{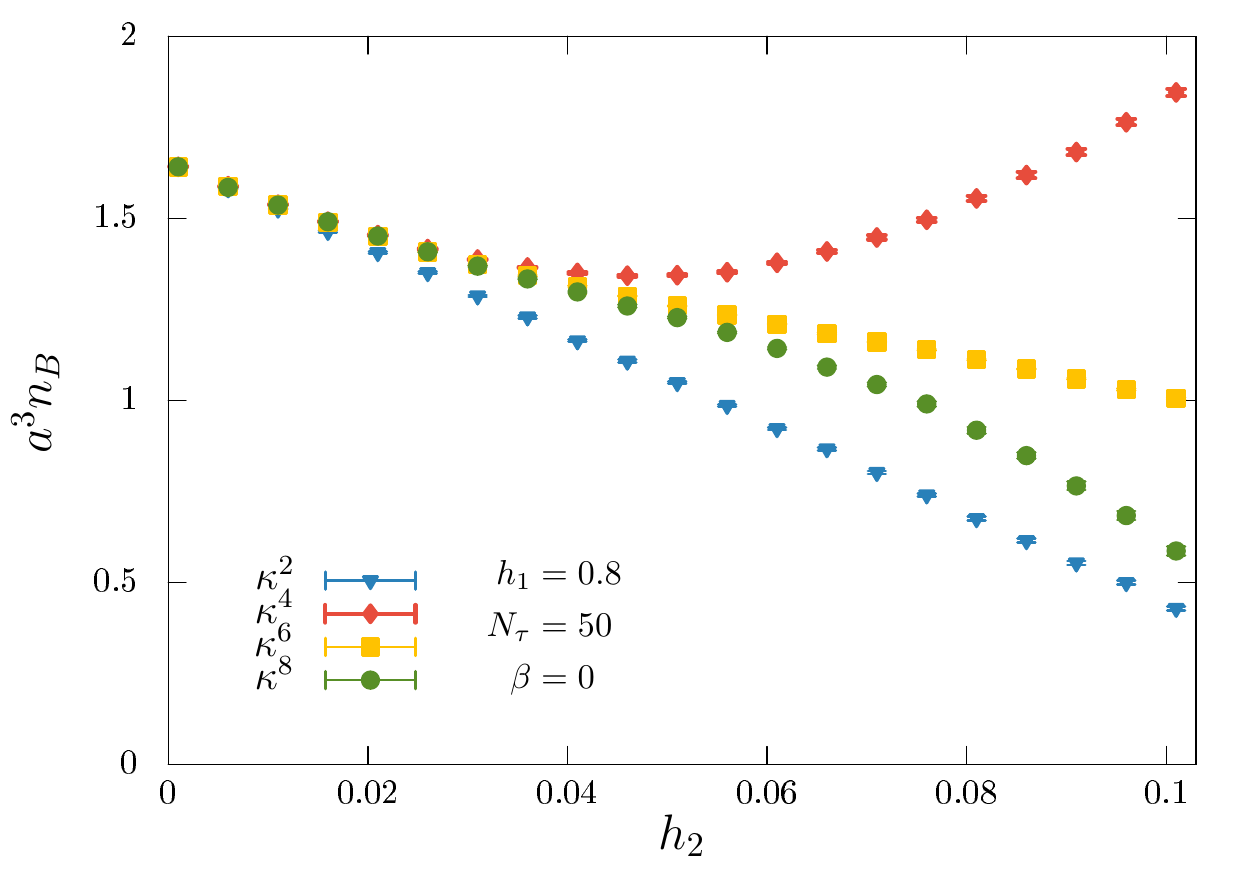}
\includegraphics[width=0.5\textwidth]{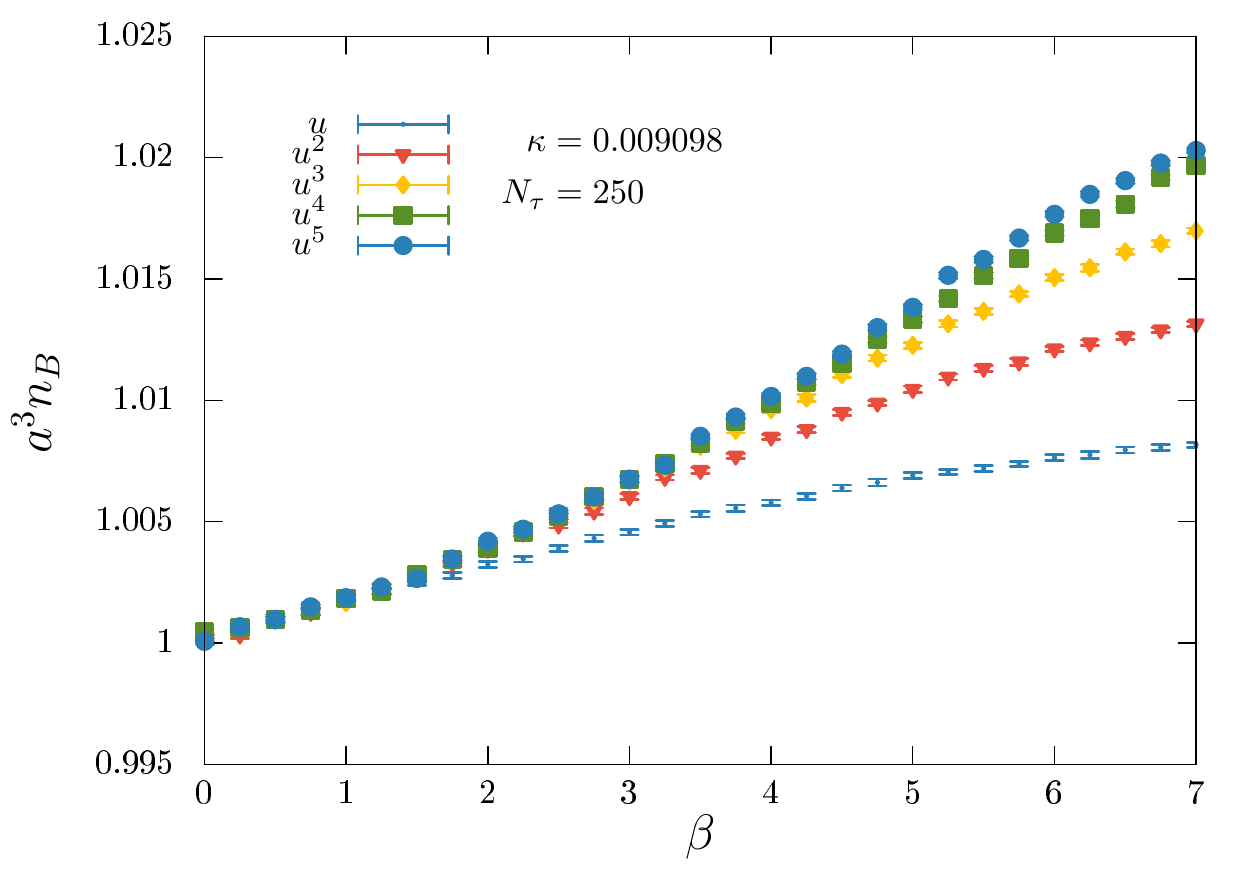}
}
\caption[]{Left: Convergence of the baryon density as a function of $h_2$, computed with 
effective actions of different orders in the hopping expansion. Right: Convergence in $u$.}
\label{fig:conv}
\end{figure}
We simulate our effective theory as described in \cite{densek4} by cross checking complex Langevin simulations
with simulations using standard Metropolis updates and reweighting. 
Our first task is to assess the range of validity of our new action. One expects 
the additional orders in $\kappa$ to extend the convergence region, within which
the description of thermodynamic functions by the effective action is reliable. We test this
by computing the baryon number density at fixed values of the coupling $h_1$ and $N_{\tau}$.
Varying $\kappa$ then allows us to assess the convergence of the expansion of the kinetic quark determinant. 
\fig\ref{fig:conv} (left) shows the results obtained with effective actions of increasing order in $\kappa$. One observes
clearly how two adjacent orders stay together for larger values of $h_2(\kappa)$ as the order is increased,
thus extending the range where our effective action is reliable.  \fig\ref{fig:conv} (right) shows
the same exercise for the largest $\kappa$ considered in this work, this time increasing the orders
of the character expansion. We observe good convergence up to $\beta\sim 6$, which is a 
sufficiently weak coupling to allow for continuum extrapolations. It is interesting to note that the
convergence properties are not determined by the size of the expansion parameters alone.
Even though the $u(\beta)$-values far exceed the $\kappa$-values employed in the 
figures, convergence in $u(\beta)$ appears to be faster.
The gain in convergence region by the additional orders in the effective action 
can be exploited to study the systematics of our effective theory.

\subsection{Continuum approach}

\begin{figure}[t]
\centerline{
\includegraphics[width=0.5\textwidth]{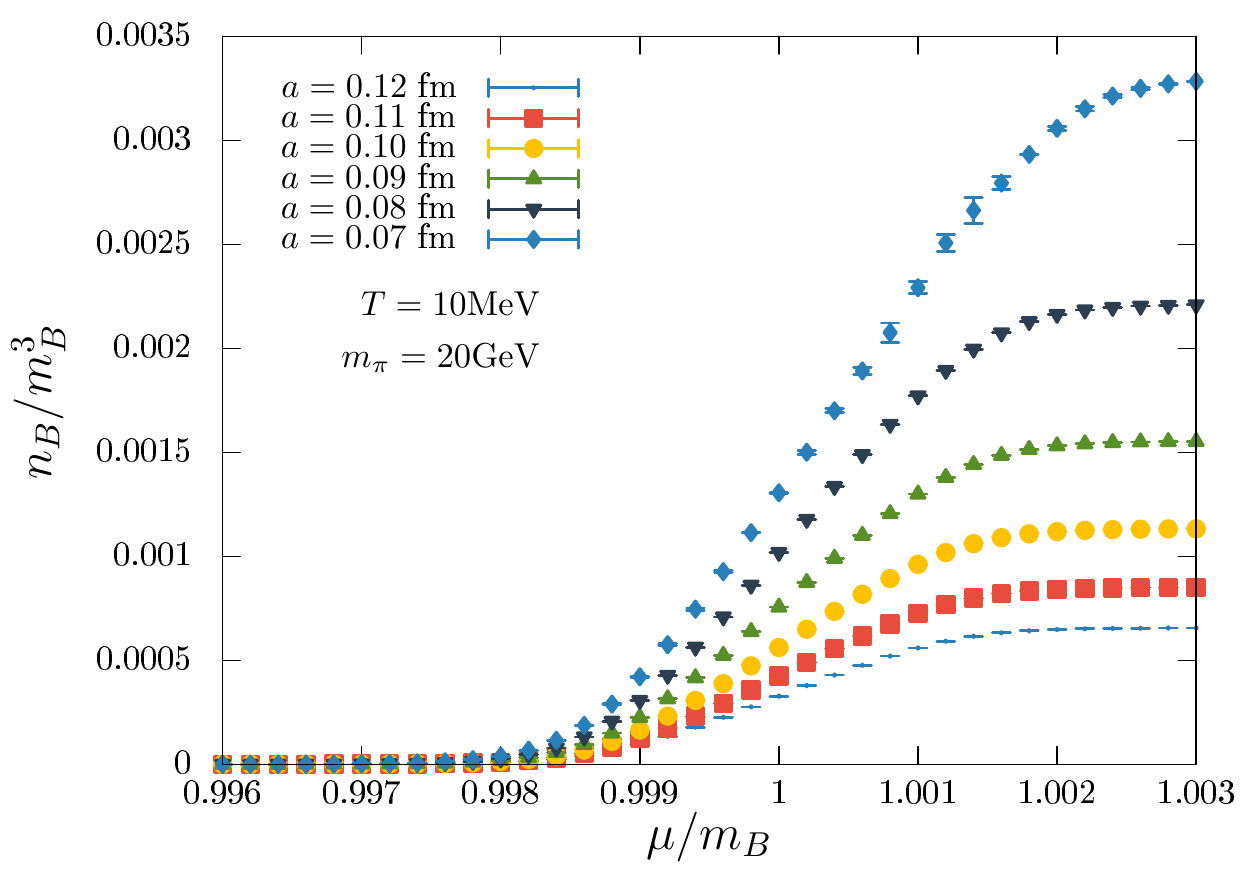}
\includegraphics[width=0.5\textwidth]{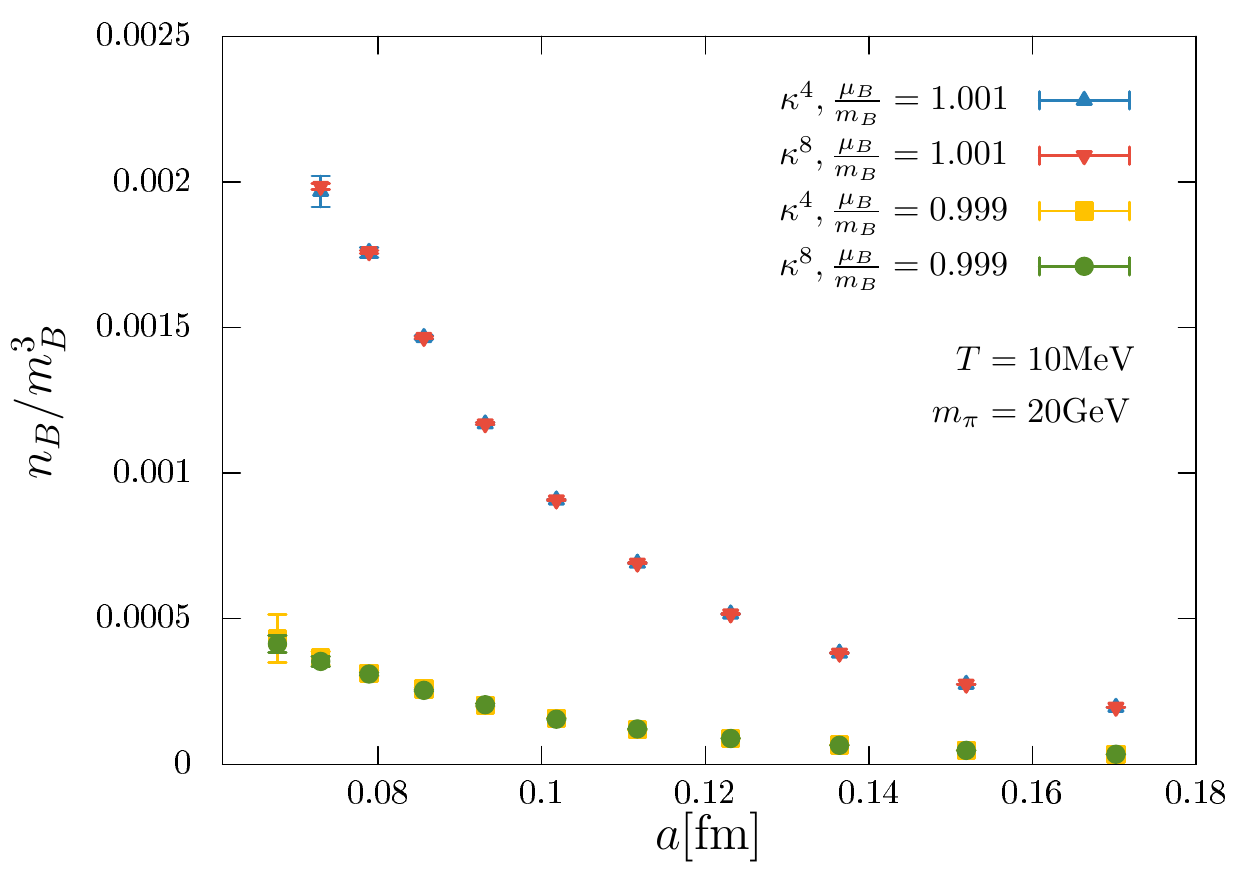}
}
\caption[]{Continuum approach of the baryon number.}
\label{fig:cont}
\end{figure}
An important question for any lattice investigation concerns the continuum limit. 
\fig\ref{fig:cont} (left) shows the baryon number as a function of chemical potential and 
highlights a severe issue of lattice QCD at finite baryon density,
irrespective of the sign problem or the accuracy of effective actions: 
cut-off effects at finite density cause not only quantitative systematic errors, 
but alter the qualitative behaviour of the system.
Because of the finite number of lattice sites available, 
the Pauli principle leads to a saturation density of $n_B^{sat}=2N_f$ baryons
per site, which does not exist in the continuum. Once lattice saturation is reached, a further 
increase of chemical potential makes no sense. Thus lattices have to be made finer 
before higher densities can be addressed. On finer lattices the saturation density in physical units 
grows and in the continuum limit moves to infinity.
This lattice artefact starts to make itself felt already 
quite early, as is also apparent in the numerical behaviour of the Polyakov loop \cite{silver} 
and related to the half-filling symmetry of the static action \cite{rindli}.

The difficulty is also reflected in \fig\ref{fig:cont} (right), 
where the slopes 
of the continuum approach rapidly increase 
with growing chemical potential, such that a continuum extrapolation is increasingly difficult to control. 
The figure shows results from our previous simulations
obtained with the $\kappa^4$ action at two values of $\mu>\mu_c$, i.e.~beyond the 
nuclear onset transition, and compares it with the new $\kappa^8$ action.
The baryon density 
just about reaches the domain with leading cut-off effects linear in $a$, which are expected
for standard Wilson fermions. In this context it should prove particularly valuable 
to work with an improved action with
${\cal{O}}(a)$ lattice corrections removed.
For still finer lattices the data points break away from this behaviour, signalling the 
limit of validity of our finite series. We conclude that the hopping expansion is systematic and controlled,
with additional orders in the action allowing to access finer lattices, but the progress is very slow. 
For sufficiently heavy masses and not too high densities, we have thus attempted a continuum 
extrapolation \cite{densek4}, which we reproduce in \fig\ref{fig:press} below. 

Another comment regarding our extrapolation is
in order. The region with linear cut-off dependence starts at $a\sim 0.1$ fm, resulting in $am_\pi\sim 10$. 
Thus our lattices are too coarse to resolve the structure of hadrons, which effectively appear as point 
particles, and one might wonder how this could possibly be consistent with continuum physics.
Indeed, the hadronic mass values
corresponding to the formulae (\ref{eq:meson_mass}, \ref{eq:baryon_mass}) are afflicted by large cut-off effects 
and do not represent the true mass values in the continuum. 
In principle this could be repaired by the methods of heavy quark effective 
theory \cite{heitger}, which however is beyond our present interest.
On the other hand, the nuclear physics in this parameter region is effectively governed by the interactions between
baryons and not within baryons. Moreover, in the case of very heavy mesons the Yukawa potential between
nuclei is extremely short ranged, i.e. in that limit the nucleons really do interact as point-like particles as
in our setup.  Thus our extrapolation should reflect continuum physics, though we only roughly know the
hadron masses this limit corresponds to.

\subsection{Mass dependence}

\begin{figure}[t]
\centerline{
\includegraphics[width=0.45\textwidth]{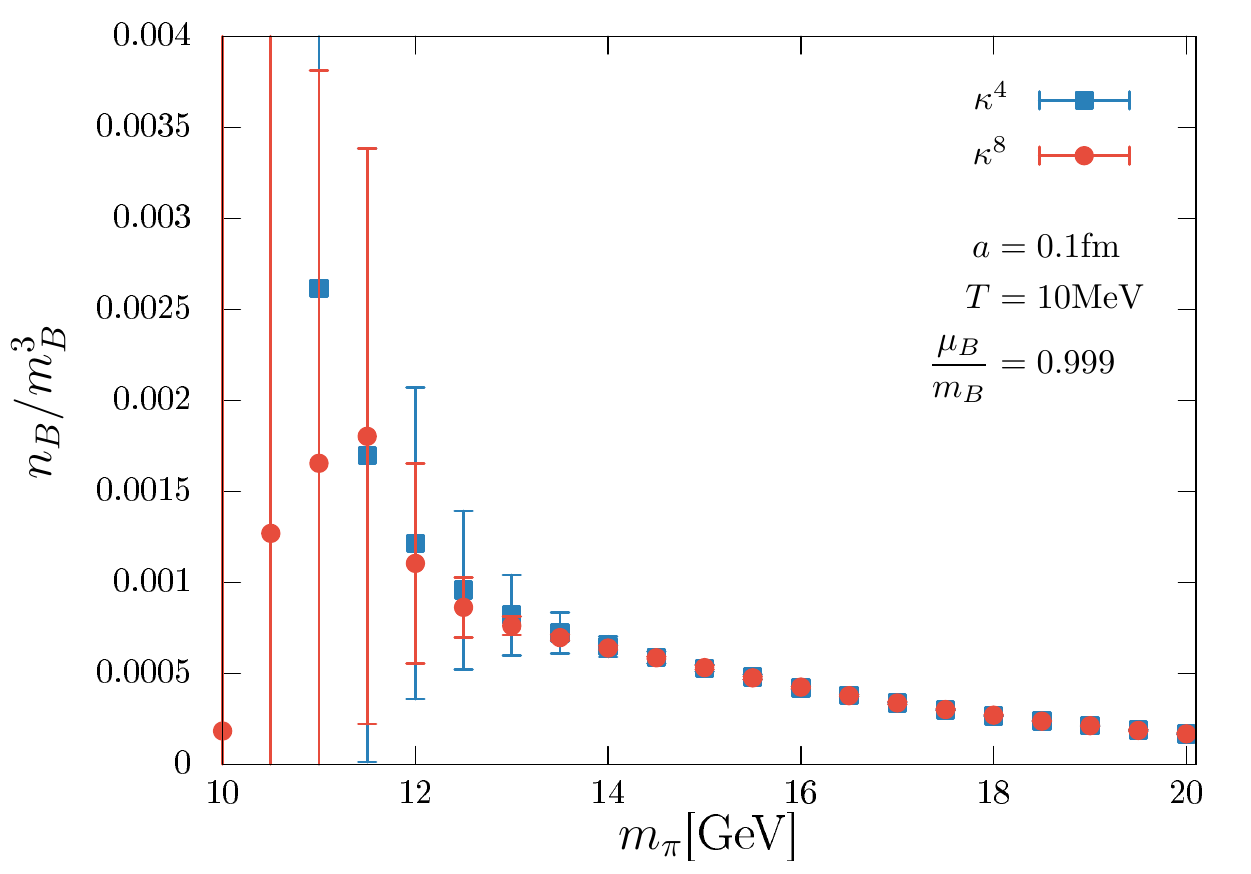}
\includegraphics[width=0.45\textwidth]{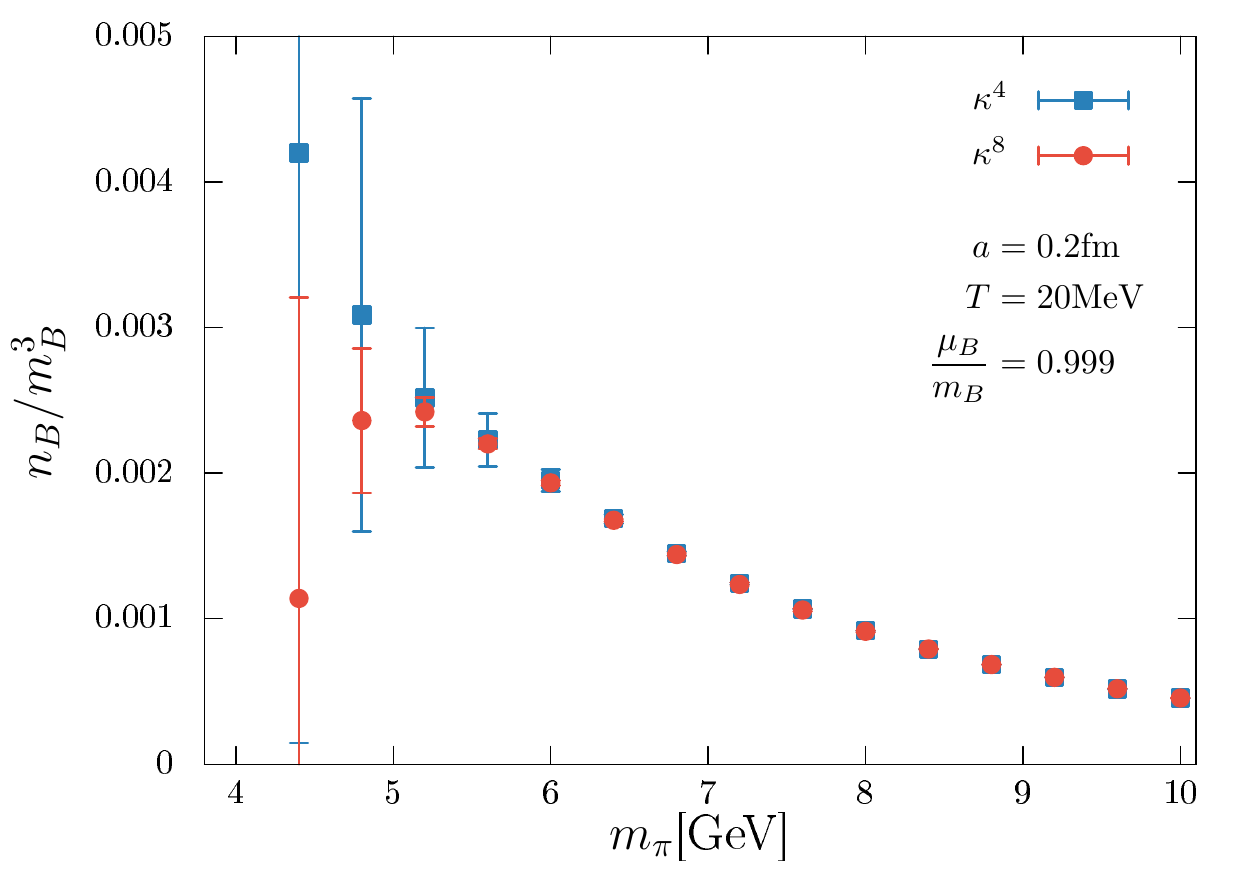}
}
\caption[]{Baryon number density as a function of pion mass.}
\label{fig:mass}
\end{figure}

A second way to benefit from the additional orders in the hopping expansion is to 
keep the lattice spacing fixed and study smaller masses. This is shown in \fig\ref{fig:mass}
for two different lattice spacings. The error bars in these plots are systematic and give 
the difference between results obtained by the action to the highest two orders in the hopping
expansion. Growing error bars thus indicate the loss of good convergence and control.
Again, this behaviour is in complete accord with qualitative expectations, with increasing orders
in the hopping expansion making smaller quark masses accessible and coarser lattices 
allowing for lighter quarks. However, these results also illustrate the fundamental difficulties and
limitations of an effective theory based on the hopping expansion.  While the systematics 
appears to be controllable and reliably tell us about its breakdown, the gain in mass range 
per additional order in the hopping expansion appears to be too small to envisage an extension 
to the physical quark masses of QCD at present. 

\section{Linked cluster expansion for the effective theory} \label{sec:linked_cluster}

So far we have derived an effective, three-dimensional theory for QCD thermodynamics
in the cold heavy mass regime and used it for numerical
simulations. However, we have observed in previous publications \cite{Bergner:2013qaa,densek4}
that the couplings of the effective theory are sufficiently small to suggest a perturbative 
calculation of thermodynamic functions. 
In this section we develop a systematic expansion scheme by applying
the linked cluster expansion, well-known from spin models \cite{wortis}. In order to apply it, we change
variables to Polyakov loops $L(\vec{x})$, which are complex numbers and resemble continuous spins. 
For the transformation of the measure, see \cite{ym}. The rational expressions of Wilson lines in the effective
action can be converted using the generating functional given in appendix \ref{app:genFunc}.

\subsection{General framework}

We begin by summarising the basic features of the linked cluster expansion,
for a more thorough review, see \cite{wortis}. 
Consider an $N$-component scalar field with a 2-point coupling, which may also extend
over larger distances than nearest neighbour,
\begin{equation}
  Z = \int \mathcal{D} \phi \; e^{-S_0[\phi] + \frac{1}{2} \sum_{x,y} \sum_{i,j} \phi_i(x) v_{ij}(x,y) \phi_j(y)}\;.
\end{equation}
All information on the interaction is encoded in $v_{ij}(x,y)$, which we
assume to be small. We will see later that in our case $v \sim \kappa^2$. Our goal is to study thermodynamic quantities, so we are 
interested in the free energy rather than the partition function, 
\begin{equation}
  \mathcal{W} = - \ln Z\;.
\end{equation}
The linked cluster expansion is thus
defined by the series expansion of $\mathcal{W}$ in powers of the coupling, 
\begin{equation}
  \mathcal{W}[v] = \bigg[ \exp\bigg( \frac{1}{2} \sum_{x,y} \sum_{i,j} v_{ij}(x,y) 
  \frac{\delta}{\delta \tilde{v}_{ij}(x,y)} \bigg) \bigg] \mathcal{W}[\tilde{v}] \Bigg|_{\tilde{v}=0}\;.
\end{equation}
A systematic way of taking the derivatives with respect to the coupling is by
introducing source terms to define the generating functionals 
\begin{align}
  Z[J] &= \int \mathcal{D} \phi \, e^{-S[\phi] + \sum_x \sum_i J_i(x) \phi_i(x)}\;,\\
  \mathcal{W}[J] &= - \ln Z[J]\;.
\end{align}
A derivative in $v$ is now replaced by
\begin{equation}
  \frac{\delta \mathcal{W}}{\delta v_{ij}(x,y)} = \frac{\delta^2 \mathcal{W}}{\delta J_i(x) \delta J_j(y)}
   + \frac{\delta \mathcal{W}}{\delta J_i(x)}\frac{\delta \mathcal{W}}{\delta J_j(y)}\;.
\end{equation}
The derivatives of the free energy with respect to the sources 
are the cumulants, or connected $n$-point functions, e.g.
\begin{equation}
  \frac{\delta^2 \mathcal{W}}{\delta J_i(x) \delta J_j(y)} = \langle \phi_i(x) \phi_j(y) \rangle - \langle \phi_i(x) \rangle
    \langle \phi_j(y) \rangle\;.
\end{equation}
Finally, setting the interaction to zero means that the cumulants only give a contribution for 
fields on the same site, which we use to define the $n$-leg expressions $M_{i_1\ldots i_n}$, 
\begin{subequations}
  \begin{align}
    M_i(x) &= \frac{\delta \mathcal{W}}{\delta J_i(x)} \bigg|_{v=0} = \langle \phi_i(x) \rangle \,,\\
    M_{ij}(x) &= \frac{\delta \mathcal{W}}{\delta J_i(x) \delta J_j(y)} \bigg|_{\mathrlap{v=0}} \quad \delta(x-y) = 
      \langle \phi_i(x) \phi_j(x) \rangle - \langle \phi_i(x) \rangle \langle \phi_j(x) \rangle \,, \\
    &\vdots  \nonumber\\
    M_{i_1\dots i_n} &= \frac{\delta \mathcal{W}}{\delta J_{i_1}(x_1) \cdots \delta J_{i_n}(x_n)} \bigg|_{\mathrlap{v=0}} \quad 
      \delta(x_1-x_2) \cdots \delta(x_{n-1} - x_n)\;.
  \end{align}
\end{subequations}
Thus we get the series expansion of $\mathcal{W}$,
\begin{align}
  \mathcal{W} = \mathcal{W}_0 &+ \frac{1}{2}\sum_{x,y} \sum_{i,j} M_i(x) v_{ij}(x,y) M_j(y) \nonumber \\
  &+ \frac{1}{2} \sum_{i,j,k} \sum_{x,y,z} M_i(x) v_{ij}(x,y) M_{jk}(y) v_{kl}(y,z) M_l(z) \nonumber \\
  &+ \frac{1}{4} \sum_{i,j} \sum_{x,y} M_{ij}(x) v_{ik}(x,y)v_{jl}(x,y) M_{kl}(y) + \mathcal{O}(v^3)\;.
\label{eq:wseries}
\end{align}

\subsection{Graphs and embeddings}

The last expression suggests a graphical notation where the $M$'s are $n$-legged nodes and the 
$v$'s are bonds connecting them. It is
apparent that the order of a node is determined by the number of bonds entering it, e.g.,
\begin{equation}
  \begin{midtikzpicture}
    \node[dot] (center) {};
    \foreach \i in {60,120,...,360}
      \draw[thin] (center) -- ({\i+30}:.5);
  \end{midtikzpicture}
  \: = \: M_{ijklmn}(x)\sim v^6\;.
\end{equation}
With this notation we can express the expansion (\ref{eq:wseries}) by graphs,
\begin{equation}
  \mathcal{W} \:=\:
  \midtikz \node[smalldot] {};
  + \textstyle\frac{1}{2}  \: \begin{midtikzpicture}
    \node[smalldot] (n1) {};
    \node[smalldot] (n2) [below=.5 of n1] {};
    \draw (n1) -- (n2);
  \end{midtikzpicture}
  + \textstyle\frac{1}{2}  \: \begin{midtikzpicture}
    \node[smalldot] (n1) {};
    \node[smalldot] (n2) at ([shift={(60:{1/sqrt(3)})}] n1) {};
    \node[smalldot] (n3) at ([shift={(300:{1/sqrt(3)})}] n2) {};
    \draw (n1) -- (n2);
    \draw (n2) -- (n3);
  \end{midtikzpicture}
  + \textstyle\frac{1}{4} \: \begin{midtikzpicture}
    \node[smalldot] (n1) {};
    \node[smalldot] (n2) [below=.5 of n1] {}
      edge[bend left=45] (n1) edge[bend right=45] (n1);
  \end{midtikzpicture}
  + \mathcal{O}(v^3)\;.
\end{equation}
The prefactors give the symmetry factor of a graph, which is the inverse number of 
ways one can label the bonds and the nodes while keeping the same mathematical expression 
(i.e.~connecting the same pairs). The expansion of $\mathcal{W}$ to some power $n$ of $v$ 
now requires computing all graphs with $n$ bonds. 

So far the graphical notation does not contain information about the spatial dependence of a graph. 
For a translationally invariant theory
all spatial dependence can be summed up in a single embedding number, 
which counts the number of ways to put a graph on the lattice.
It depends on the type and lattice distance of the interaction,
as well as the dimension and geometry of the lattice we are working on. For example the $v^3$ term
\begin{center}
  \begin{tikzpicture}
    \node[smalldot] (n1) {};
    \node[smalldot] (n2) at ([shift={(60:{1/sqrt(3)})}] n1) {};
    \node[smalldot] (n3) at ([shift={(300:{1/sqrt(3)})}] n2) {};
    \draw (n1) -- (n2);
    \draw (n2) -- (n3);
    \draw (n3) -- (n1);
  \end{tikzpicture}
\end{center}
cannot be put on a square lattice, and thus its embedding number is 0, while on a triangular lattice
 it would be non-zero. A quick summary of the lowest order
graphs with symmetry factors and embeddings is given in table~\ref{tab:graphs}.
\begin{table}
\begin{center}
{\tabulinesep = 1.2mm
\begin{tabu}{|ccc||ccc|} \hline
  \textbf{Graph} & \textbf{Symmetries} & \textbf{Embeddings} &
  \textbf{Graph} & \textbf{Symmetries} & \textbf{Embeddings} \\ \hline
  \begin{tikzpicture}[scale=0.5, every node/.style={scale=0.75}]
    \node[smalldot] (n1) {};
    \node[smalldot] (n2) at ([shift={(0,1)}] n1) {}
      edge (n1);
  \end{tikzpicture} & 2 & $2d$ &
  \begin{tikzpicture}[scale=0.4, every node/.style={scale=0.75}]
    \node[smalldot] (n1) {};
    \node[smalldot] (n2) at ([shift={(330:1)}] n1) {}
      edge (n1);
    \node[smalldot] (n3) at ([shift={(90:1)}] n1) {}
      edge (n1);
    \node[smalldot] (n4) at ([shift={(210:1)}] n1) {}
      edge (n1);
  \end{tikzpicture} & $3!$ & $(2d)^3$ \\ \hline
  \begin{tikzpicture}[scale=0.5, every node/.style={scale=0.75}]
    \node[smalldot] (n1) {};
    \node[smalldot] (n2) at ([shift={(60:1)}] n1) {}
      edge (n1);
    \node[smalldot] (n3) at ([shift={(300:1)}] n2) {}
      edge (n2);
  \end{tikzpicture} & 2 & $(2d)^2$ &
  \begin{tikzpicture}[scale=0.5, every node/.style={scale=0.75}]
    \node[smalldot] (n1) {};
    \node[smalldot] (n2) at ([shift={(60:1)}] n1) {}
      edge (n1);
    \node[smalldot] (n3) at ([shift={(300:1)}] n2) {}
      edge (n2) edge (n1);
  \end{tikzpicture} & 6 & 0 \\ \hline
  \begin{tikzpicture}[scale=0.5, every node/.style={scale=0.75}]
    \node[smalldot] (n1) {};
    \node[smalldot] (n2) at ([shift={(0,1)}] n1) {}
      edge[bend left=45] (n1) edge[bend right=45] (n1);
  \end{tikzpicture} & 4 & $2d$ &
  \begin{tikzpicture}[scale=0.5, every node/.style={scale=0.75}]
    \node[smalldot] (n1) {};
    \node[smalldot] (n2) at ([shift={(60:1)}] n1) {}
      edge[bend left=45] (n1) edge[bend right=45] (n1);
    \node[smalldot] at ([shift={(300:1)}] n2) {}
      edge (n2);
  \end{tikzpicture} & 2 & $(2d)^2$ \\ \hline
  \begin{tikzpicture}[scale=0.5, every node/.style={scale=0.75}]
    \node[smalldot] (n1) {};
    \node[smalldot] (n2) at ([shift={(60:1)}] n1) {}
      edge (n1);
    \node[smalldot] (n3) at ([shift={(300:1)}] n2) {}
      edge (n2);
    \node[smalldot] (n4) at ([shift={(60:1)}] n3) {}
      edge (n3);
  \end{tikzpicture} & 2 & $(2d)^3$ &
  \begin{tikzpicture}[scale=0.5, every node/.style={scale=0.75}]
    \node[smalldot] (n1) {};
    \node[smalldot] at ([shift={(0,1)}] n1) {}
      edge[bend right=45] (n1) edge[bend left=45] (n1)
      edge (n1);
  \end{tikzpicture} & $2 {\scriptstyle\times} 3!$ & $2d$ \\ \hline
\begin{tikzpicture}[scale=0.5, every node/.style={scale=0.75}]
  \node[smalldot] (n1) {};
  \node[smalldot] (n2) at ([shift={(60:1)}] n1) {}
    edge (n1);
  \node[smalldot] (n3) at ([shift={(300:1)}] n2) {}
    edge (n2);
  \node[smalldot] (n4) at ([shift={(60:1)}] n3) {}
    edge (n3);
  \node[smalldot] at ([shift={(300:1)}] n4) {}
    edge (n4);
\end{tikzpicture} & 2 & $(2d)^4$ &
\begin{tikzpicture}[scale=0.4, every node/.style={scale=0.75}]
  \node[smalldot] (n1) {};
  \node[smalldot] at ([shift={(45:1)}] n1) {}
    edge (n1);
  \node[smalldot] at ([shift={(135:1)}] n1) {}
    edge (n1);
  \node[smalldot] at ([shift={(225:1)}] n1) {}
    edge (n1);
  \node[smalldot] at ([shift={(315:1)}] n1) {}
    edge (n1);
\end{tikzpicture} & $4!$ & $(2d)^4$ \\ \hline
\begin{tikzpicture}[scale=0.5, every node/.style={scale=0.75}]
  \node[smalldot] (n1) {};
  \node[smalldot] (n2) at ([shift={(60:1)}] n1) {}
    edge (n1);
  \node[smalldot] (n3) at ([shift={(300:1)}] n2) {}
    edge (n2);
  \node[smalldot] (n4) at ([shift={(60:1)}] n3) {}
    edge (n3);
  \node[smalldot] at ([shift={(0:1)}] n3) {}
    edge (n3);
\end{tikzpicture} & 2 & $(2d)^4$ &
\begin{tikzpicture}[scale=0.4, every node/.style={scale=0.75}]
  \node[smalldot] (n1) {};
  \node[smalldot] (n2) at ([shift={(1.25,0)}] n1) {}
    edge (n1);
  \node[smalldot] (n3) at ([shift={(0,1.25)}] n2) {}
    edge (n2);
  \node[smalldot] at ([shift={(0,1.25)}] n1) {}
    edge (n1) edge(n3);
\end{tikzpicture} & $8$ & $2(2d)^2 - 1$ \\ \hline
\begin{tikzpicture}[scale=0.4, every node/.style={scale=0.75}]
    \node[smalldot] (n1) {};
    \node[smalldot] (n2) at ([shift={(330:1)}] n1) {}
      edge (n1);
    \node[smalldot] (n3) at ([shift={(90:1)}] n1) {}
      edge[bend left=45] (n1) edge[bend right=45] (n1);
    \node[smalldot] (n4) at ([shift={(210:1)}] n1) {}
      edge (n1);
\end{tikzpicture} & $2 {\scriptstyle\times} 2!$ & $(2d)^3$ &
\begin{tikzpicture}[scale=0.5, every node/.style={scale=0.75}]
  \node[smalldot] (n1) {};
  \node[smalldot] (n2) at ([shift={(60:1)}] n1) {}
    edge (n1);
  \node[smalldot] (n3) at ([shift={(300:1)}] n2) {}
    edge (n2);
  \node[smalldot] (n4) at ([shift={(60:1)}] n3) {}
    edge[bend right=45] (n3) edge[bend left=45] (n3);
\end{tikzpicture} & $2$ & $(2d)^3$ \\ \hline
\begin{tikzpicture}[scale=0.5, every node/.style={scale=0.75}]
  \node[smalldot] (n1) {};
  \node[smalldot] (n2) at ([shift={(60:1)}] n1) {}
    edge (n1);
  \node[smalldot] (n3) at ([shift={(300:1)}] n2) {}
    edge[bend right=45] (n2) edge[bend left=45] (n2);
  \node[smalldot] (n4) at ([shift={(60:1)}] n3) {}
    edge (n3);
\end{tikzpicture} & $2 {\scriptstyle\times} 2!$ & $(2d)^3$ &
\begin{tikzpicture}[scale=0.5, every node/.style={scale=0.75}]
  \node[smalldot] (n1) {};
  \node[smalldot] (n2) at ([shift={(60:1)}] n1) {}
    edge (n1);
  \node[smalldot] (n3) at ([shift={(300:1)}] n2) {}
    edge (n2) edge[bend right=45] (n1) edge[bend left=45] (n1);
\end{tikzpicture} & $2 {\scriptstyle\times} 2!$ & $0$ \\ \hline
\begin{tikzpicture}[scale=0.5, every node/.style={scale=0.75}]
  \node[smalldot] (n1) {};
  \node[smalldot] (n2) at ([shift={(60:1)}] n1) {}
    edge (n1);
  \node[smalldot] at ([shift={(1,0)}] n2) {}
    edge (n2);
  \node[smalldot] (n3) at ([shift={(300:1)}] n2) {}
    edge (n2) edge (n1);
\end{tikzpicture} & $2$ & $0$ &
\begin{tikzpicture}[scale=0.5, every node/.style={scale=0.75}]
  \node[smalldot] (n1) {};
  \node[smalldot] (n2) at ([shift={(60:1)}] n1) {}
    edge [bend left=30] (n1) edge [bend right=30] (n1);
  \node[smalldot] (n3) at ([shift={(300:1)}] n2) {}
    edge[bend right=30] (n2) edge[bend left=30] (n2);
\end{tikzpicture} & $2 {\scriptstyle\times} (2!)^2$ & $(2d)^2$ \\ \hline
\begin{tikzpicture}[scale=0.5, every node/.style={scale=0.75}]
  \node[smalldot] (n1) {};
  \node[smalldot] (n2) at ([shift={(60:1)}] n1) {}
    edge (n1) edge [bend left=30] (n1) edge [bend right=30] (n1);
  \node[smalldot] (n3) at ([shift={(300:1)}] n2) {}
    edge (n2);
\end{tikzpicture} & $3!$ & $(2d)^2$ &
\begin{tikzpicture}[scale=0.5, every node/.style={scale=0.75}]
  \node[smalldot] (n1) {};
  \node[smalldot] (n2) at ([shift={(90:1)}] n1) {}
    edge [bend left=20] (n1) edge [bend right=20] (n1)
    edge [bend left=60] (n1) edge [bend right=60] (n1);
\end{tikzpicture} & $2 {\scriptstyle\times} 4!$ & $2d$ \\ \hline
\end{tabu}
}
\end{center}
\caption{Graphs up to 4 bonds together with symmetries and lattice embeddings on a $d$ dimensional square lattice}
\label{tab:graphs}
\end{table}

We are now ready to map our effective theory to $\mathcal{O}(\kappa^2)$ 
onto this computational scheme.
The partition function to this order contains a nearest neighbour interaction
between two $W_{1,1}$-terms,
\begin{equation}
  Z = \int \mathcal{D} W\, \big( \det Q_{\mathrm{stat}} \big)^{2 N_f}
  e^{-\frac{1}{2} \sum_{x,y} W_{1,1}(x) \big( 2 h_2 N_f \sum_i \delta (x + i - y) \big) W_{1,1}(y)}\;.
\end{equation}
We can thus apply the results from the linked cluster expansion for a one component field 
$\phi_1(x)=W_{1,1}(x)$ by identifying 
\beq
v_{11}(x,y)=2h_2N_f\sum_i\delta(x+i-y)\;.
\eeq
The building blocks for (\ref{eq:wseries}) are then
\begin{subequations}
\begin{align}
  \mathcal{W}_0 &= \int \mathrm{d} W(x) \, \big( \det Q_{\mathrm{stat}} \big)^{2 N_f} \;,\\
  M_i(x) &= \int \mathrm{d} W(x) \, \big( \det Q_{\mathrm{stat}} \big)^{2 N_f} W_{1,1}(x)\;, \\
  M_{ij}(x) &= \int \mathrm{d} W(x) \, \big( \det Q_{\mathrm{stat}} \big)^{2 N_f} W_{1,1}^2(x) 
  - M_i(x)M_j(x) \;.
\end{align}
\end{subequations}

\subsection{Higher order couplings}

At $\mathcal{O}(\kappa^4)$ we are confronted with 3-point couplings. 
Fortunately, introducing higher $n$-point interactions to the linked cluster expansion 
is straightforward. In our case we need a generalised partition function
\begin{equation}
  Z = \int \mathcal{D} \phi \, e^{-S_0[\phi] + \frac{1}{2} \sum v_{ij} (x,y) \phi_i(x) \phi_j(y)
  + \frac{1}{3!} \sum u_{ijk}(x,y,z) \phi_i(x) \phi_j(y) \phi_k(z) + \dots}
\end{equation}
which has a cluster expansion
\begin{align}
  \mathcal{W}[v,u] = \bigg[ &\exp\bigg( \frac{1}{2} \sum_{x,y} \sum_{i,j} v_{ij}(x,y) 
  \frac{\delta}{\delta \tilde{v}_{ij}(x,y)} \bigg)  \nonumber \\
  &\times\exp\bigg( \frac{1}{3!} \sum_{x,y,z} \sum_{i,j,k} u_{ijk}(x,y,z)
  \frac{\delta}{\delta \tilde{u}_{ijk}(x,y,z)} \bigg) \cdots
  \bigg] \mathcal{W}[\tilde{v},\tilde{u}] \Bigg|_{\substack{\tilde{v}=0\\\tilde{u}=0\\\cdots}}\;,
\end{align}
where the derivative with respect to $\tilde{u}$ is once more given by the cumulants,
\begin{align}
  \frac{\delta \mathcal{W}}{\delta u_{ijk}(x,y,z)} &= \frac{\delta^3 \mathcal{W}}{\delta J_i(x) \delta J_j(y) \delta J_k(z)}
  + \frac{\delta \mathcal{W}}{\delta J_i(x)} \frac{\delta^2 \mathcal{W}}{\delta J_j(y) \delta J_k(z)}
  + \frac{\delta \mathcal{W}}{\delta J_j(y)} \frac{\delta^2 \mathcal{W}}{\delta J_i(x) \delta J_k(z)} \nonumber \\
  &\hspace{2cm} + \frac{\delta \mathcal{W}}{\delta J_k(z)} \frac{\delta^2 \mathcal{W}}{\delta J_i(x) \delta J_j(y)}
  + \frac{\delta \mathcal{W}}{\delta J_i(x)} \frac{\delta \mathcal{W}}{\delta J_j(y)} \frac{\delta \mathcal{W}}{\delta J_k(z)}\;.
\end{align}
The geometry of the interaction term is contained in $u_{ijk}(x,y,z)$. 
For example if we take $\phi$ as a
two-component field, $\phi = \big\{ W_{1,1}, W_{2,1} \big\}$, the three-point $\mathcal{O}(\kappa^4)$ term has 
an interaction tensor
\begin{subequations}
\begin{align}
  u_{1jk}(x,y,z) &= 2 h_2^2 N_f \sum_{\hat{a},\hat{b}} 
  \begin{pmatrix} 0 & \delta(z + \hat{a} - x) \delta(z + \hat{b} - y) \\
    \delta(y + \hat{a} - x) \delta(y + \hat{b} - z) & 0 \end{pmatrix}_{jk} \;,\\
  u_{2jk}(x,y,z) &= 2 h_2^2 N_f \sum_{\hat{a},\hat{b}} 
  \begin{pmatrix} \delta(x + \hat{a} - y) \delta(x + \hat{b} - z) & 0 \\
    0 & \makebox[4cm][c]{0} \end{pmatrix}_{jk}\;,
\end{align}
\end{subequations}
corresponding to a wedge.
In this case the linked cluster expansion of $\mathcal{W}$ is the sum of all diagrams which can be 
made out of these two components,
\begin{equation}
  W \:=\:
  \midtikz \node[dot] {};
  + \textstyle\frac{1}{2}  \: \begin{midtikzpicture}
    \node[dot] (n1) {};
    \node[dot] (n2) [below=.5 of n1] {};
    \draw[thick] (n1) -- (n2);
  \end{midtikzpicture}
  + \textstyle\frac{1}{2}  \: \begin{midtikzpicture}
    \node[dot] (n1) {};
    \node[dot] (n2) at ([shift={(60:{1/sqrt(3)})}] n1) {};
    \node[dot] (n3) at ([shift={(300:{1/sqrt(3)})}] n2) {};
    \draw[thick] (n1) -- (n2);
    \draw[thick] (n2) -- (n3);
  \end{midtikzpicture}
  + \textstyle\frac{1}{4} \: \begin{midtikzpicture}
    \node[dot] (n1) {};
    \node[dot] (n2) [below=.5 of n1] {}
      edge[thick,bend left=45] (n1) edge[thick,bend right=45] (n1);
  \end{midtikzpicture}
  + \textstyle\frac{1}{2}  \: \begin{midtikzpicture}
    \node[dot,blue] (n1) {};
    \node[dot,blue] (n2) at ([shift={(60:{1/sqrt(3)})}] n1) {};
    \node[dot,blue] (n3) at ([shift={(300:{1/sqrt(3)})}] n2) {};
    \draw[directed,blue,thick] (n2) -- (n1);
    \draw[directed,blue,thick] (n2) -- (n3);
  \end{midtikzpicture}
  + \textstyle\frac{1}{2} \: \begin{midtikzpicture}
    \node[dot,blue] (n1) {};
    \node[dot,blue] (n2) [below=.5 of n1] {};
    \draw[directed,blue,thick] (n1) .. controls +(-.15,-.15) and +(-.15,.15) .. (n2);
    \draw[directed,blue,thick] (n1) .. controls +(.15,-.15) and +(.15,.15) .. (n2);
  \end{midtikzpicture}
  + \textstyle\frac{1}{2} \: \begin{midtikzpicture}
    \node[dot,red!80!black] (n1) {};
    \node[dot,red!80!black] (n2) [below=.5 of n1] {}
      edge[thick,bend left=45,red!80!black] (n1) edge[thick,bend right=45,red!80!black] (n1);
  \end{midtikzpicture}
  + \mathcal{O}(v^3)\;,
\end{equation}
where the three new diagrams come from the 2-point and 3-point terms in the $\kappa^4$-action. Note that 
directions are necessary 
to distinguish a node
$W_{2,1}$ from $W_{1,1}^2$. This also changes the symmetry factor. 
It is thus possible to
compute all graphs from combining elements up to a certain order, carefully calculating 
symmetry factors as one proceeds to higher and higher orders.

Alternatively, and as an independent check, 
one can use the idea of embedding graphs from the effective action onto
the basic graph topologies of the cluster expansion. 
As an example, consider the square graph
\begin{equation}
  \begin{tikzpicture}
    \node[smalldot] (n1) {};
    \node[smalldot] (n2) at (0.75,0) {};
    \node[smalldot] (n3) at (0.75,0.75) {};
    \node[smalldot] (n4) at (0,0.75) {};
    \draw (n1) -- (n2) -- (n3) -- (n4) -- (n1);
  \end{tikzpicture} \hspace{.5cm} \text{symmetry: } 8\;.
\end{equation}
The following $\mathcal{O}(v^4)$ terms can be embedded on it,
\begin{equation}
  \begin{tikzpicture}[baseline=-.5ex]
    \node[dot] (n1) {};
    \node[dot] (n2) [above=.35 of n1] {};
    \draw[thick] (n1) -- (n2);
  \end{tikzpicture}, \hspace{1cm}
  \begin{tikzpicture}[baseline=-.5ex]
    \node[dot,blue] (n1) {};
    \node[dot,blue] (n2) at ([shift={(60:{1/sqrt(3)})}] n1) {};
    \node[dot,blue] (n3) at ([shift={(300:{1/sqrt(3)})}] n2) {};
    \draw[directed,blue,thick] (n2) -- (n1);
    \draw[directed,blue,thick] (n2) -- (n3);
  \end{tikzpicture}, \hspace{1cm}
  \begin{tikzpicture}[baseline=-.5ex]
    \node[dot,ForestGreen] (n1) {};
    \node[dot,ForestGreen] (n2) at ([shift={(60:{1/sqrt(3)})}] n1) {};
    \node[dot,ForestGreen] (n3) at ([shift={(300:{1/sqrt(3)})}] n2) {};
    \node[dot,ForestGreen] (n4) at ([shift={(60:{1/sqrt(3)})}] n3) {};
    \draw[directed,ForestGreen,thick] (n2) -- (n1);
    \draw[ForestGreen,thick] (n2) -- (n3);
    \draw[directed,ForestGreen,thick] (n3) -- (n4);
  \end{tikzpicture}, \hspace{1cm}
  \begin{tikzpicture}[baseline=-.5ex]
    \node[dot,BurntOrange] (n1) {};
    \node[dot,BurntOrange] (n2) at ([shift={(60:{1/sqrt(3)})}] n1) {};
    \node[dot,BurntOrange] (n3) at ([shift={(300:{1/sqrt(3)})}] n2) {};
    \node[dot,BurntOrange] (n4) at ([shift={(60:{1/sqrt(3)})}] n3) {};
    \node[dot,BurntOrange] (n5) at ([shift={(300:{1/sqrt(3)})}] n4) {};
    \draw[directed,BurntOrange,thick] (n2) -- (n1);
    \draw[BurntOrange,thick] (n2) -- (n3);
    \draw[BurntOrange,thick] (n3) -- (n4);
    \draw[directed,BurntOrange,thick] (n4) -- (n5);
  \end{tikzpicture}\;,
\end{equation}
with an embedding number, that counts the number of ways this is possible.
\begin{center}
{\tabulinesep = 1.2mm
\begin{tabu}{|ccc|} \hline
  \textbf{Graph} & \textbf{Embeddings} & \textbf{Symmetry} \\ \hline
  \begin{tikzpicture}
    \node[dot] (n1) {};
    \node[dot] (n2) at (0.75,0) {};
    \node[dot] (n3) at (0.75,0.75) {};
    \node[dot] (n4) at (0,0.75) {};
    \draw[thick] (n1) -- (n2) -- (n3) -- (n4) -- (n1);
  \end{tikzpicture} & 1 & 8 \\ \hline
  \begin{tikzpicture}
    \node[dot] (n1) {};
    \node[dot] (n2) at (0.75,0) {};
    \fill[blue] (n2.base) (n2.south east) arc (-45:135:3.2pt);
    \node[dot,fill=blue] (n3) at (0.75,0.75) {};
    \node[dot] (n4) at (0,0.75) {};
    \fill[blue] (n4.base) (n4.south east) arc (-45:135:3.2pt);
    \draw[thick] (n1) -- (n2);
    \draw[thick] (n1) -- (n4);
    \draw[directed,blue,thick] (n3) -- (n2);
    \draw[directed,blue,thick] (n3) -- (n4);
  \end{tikzpicture} & 4 & 2 \\ \hline
  \begin{tikzpicture}
    \node[dot,fill=blue] (n1) {};
    \node[dot,fill=blue] (n2) at (0.75,0) {};
    \node[dot,fill=blue] (n3) at (0.75,0.75) {};
    \node[dot,fill=blue] (n4) at (0,0.75) {};
    \draw[directed,blue,thick] (n1) -- (n2);
    \draw[directed,blue,thick] (n1) -- (n4);
    \draw[directed,blue,thick] (n3) -- (n2);
    \draw[directed,blue,thick] (n3) -- (n4);
  \end{tikzpicture} & 2 & 4 \\ \hline
  \begin{tikzpicture}
    \node[dot] (n1) {};
    \node[dot,ForestGreen] (n2) at (0.75,0) {};
    \fill[ForestGreen] (n1.base) (n1.south west) arc (-135:45:3.2pt);
    \node[dot,ForestGreen] (n3) at (0.75,0.75) {};
    \node[dot] (n4) at (0,0.75) {};
    \fill[ForestGreen] (n4.base) (n4.south east) arc (-45:135:3.2pt);
    \draw[thick] (n1) -- (n4);
    \draw[directed,ForestGreen,thick] (n2) -- (n1);
    \draw[directed,ForestGreen,thick] (n3) -- (n4);
    \draw[ForestGreen,thick] (n2) -- (n3);
  \end{tikzpicture} & 4 & 2 \\ \hline
  \begin{tikzpicture}
    \node[dot,fill=BurntOrange] (n1) {};
    \node[dot,fill=BurntOrange] (n2) at (0.75,0) {};
    \node[dot,fill=BurntOrange] (n3) at (0.75,0.75) {};
    \node[dot,fill=BurntOrange] (n4) at (0,0.75) {};
    \draw[directed,BurntOrange,thick] (n4) -- (n3);
    \draw[directed,BurntOrange,thick] (n2) -- (n3);
    \draw[BurntOrange,thick] (n1) -- (n2);
    \draw[BurntOrange,thick] (n1) -- (n4);
  \end{tikzpicture} & 4 & 2 \\ \hline
\end{tabu}}
\end{center}
This will modify the symmetry factor to be
\begin{equation}
  \frac{\text{\# of unique embeddings}}{\text{graph symmetry factor}}\;.
\end{equation}
Thus we can systematically get the linked cluster expansion for our full effective theory, 
by writing down all topologically distinct diagrams and
then embed our effective theory terms onto the graphs. The result of this endeavour is too lengthy to include in this publication,
but the interested reader may obtain the full result by contacting the authors.

\subsection{Results} \label{sec:cle_results}

\begin{figure}[t]
\centerline{
\includegraphics[width=0.5\textwidth]{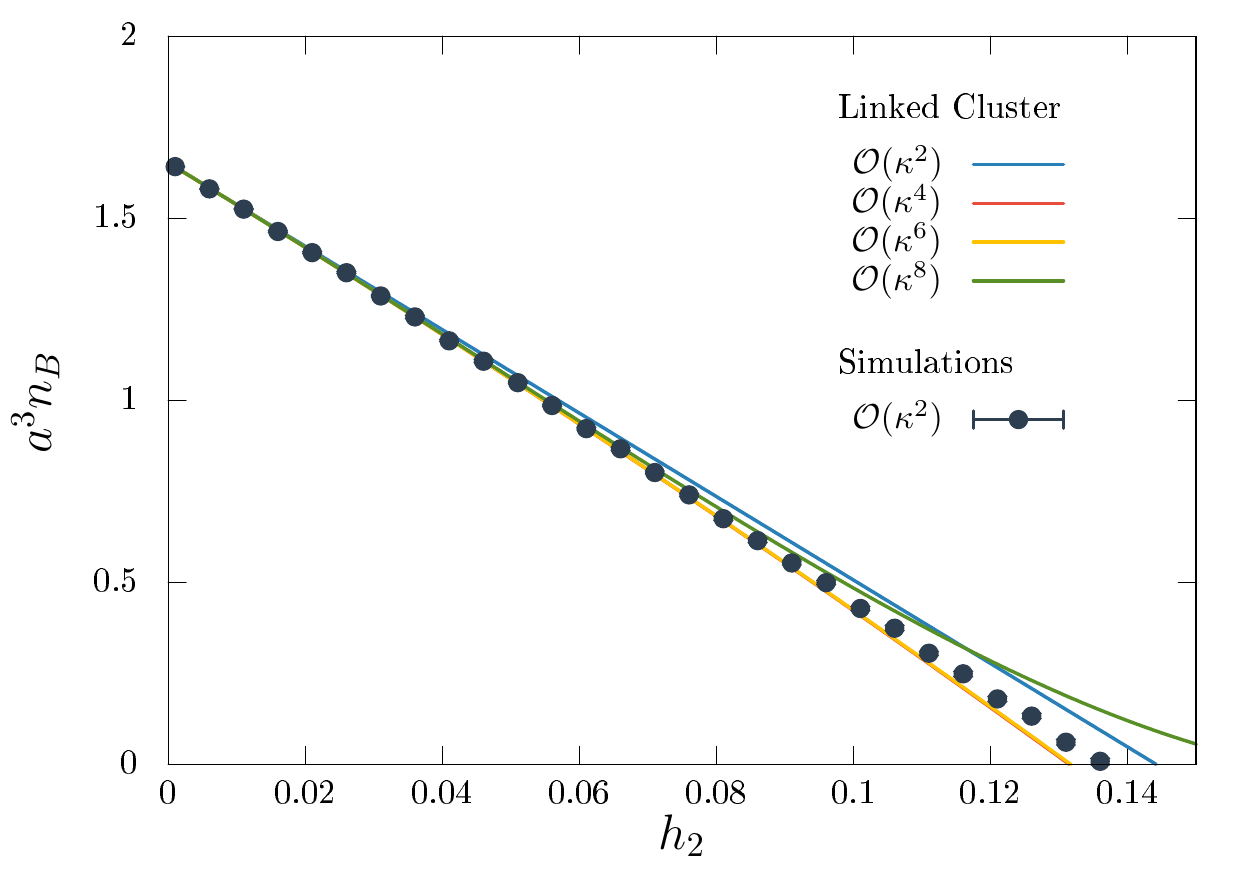}
\includegraphics[width=0.5\textwidth]{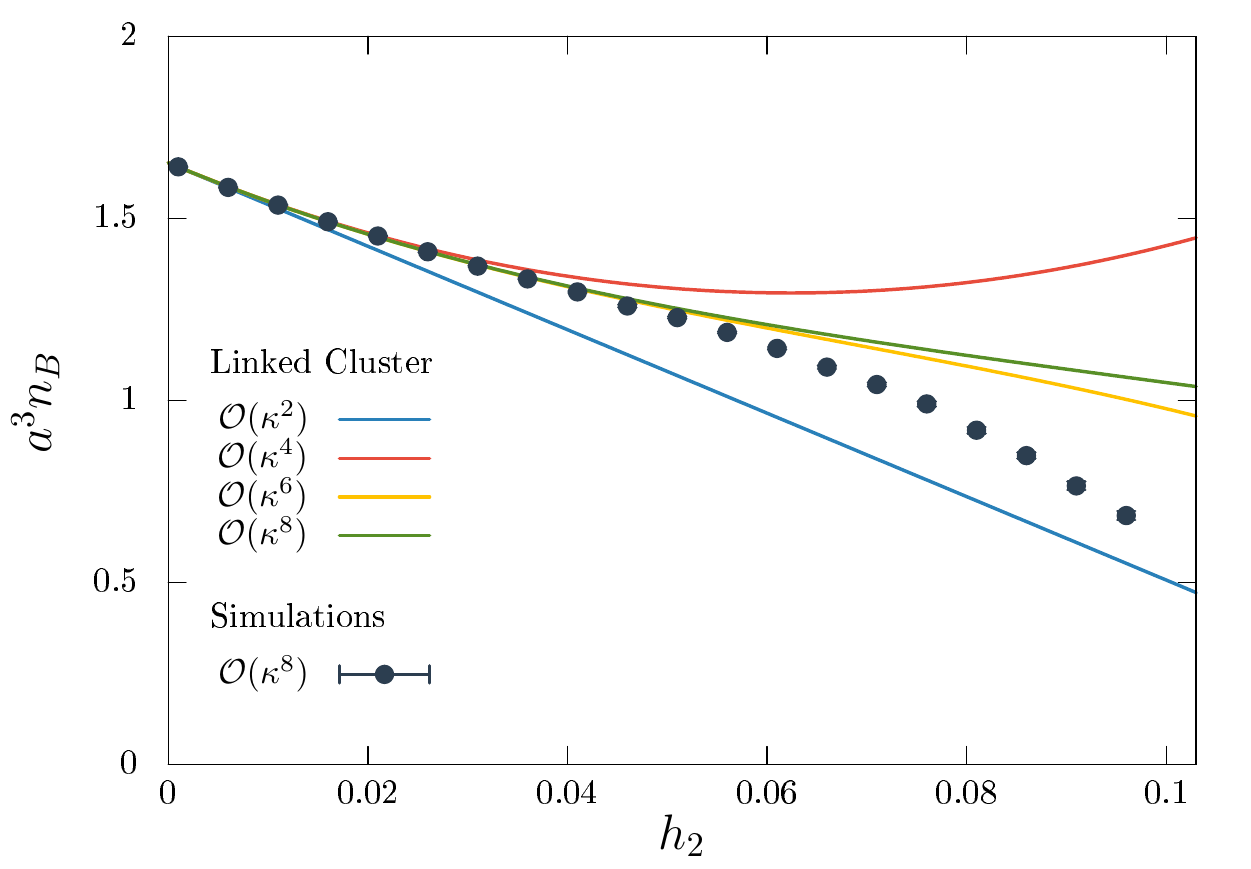}
}
\caption[]{Linked cluster expansion of the $\kappa^2$ (left) and $\kappa^8$ (right) actions for
$\beta=0$ and $h_1=0.8$.}
\label{fig:an_conv}
\end{figure}

We are now in a position to evaluate thermodynamical functions completely analytically, 
presently we have computed through order $\kappa^8$. \fig \ref{fig:an_conv} shows the evaluation
of the $\kappa^2$-action (left) and the $\kappa^8$-action (right) in various orders of the linked 
cluster expansion in comparison with the numerical evaluation, again in the 
strong coupling limit $\beta=0$. It is interesting to note the 
faster convergence for the lower order action, as one might expect. For the higher order action,
the linked cluster expansion first has to ``catch up'' to the order of the action before it can
start resumming its contributions. Comparing with \fig\ref{fig:conv} we observe that, to the orders
computed, the linked cluster expansion converges roughly as far as the derivation of the 
effective theory. If we allow for 10\% deviation 
between different orders and from the full result, the combined calculation is valid
up to $h_2\sim 0.08$.

\begin{figure}[t]
\centerline{
\includegraphics[width=0.5\textwidth]{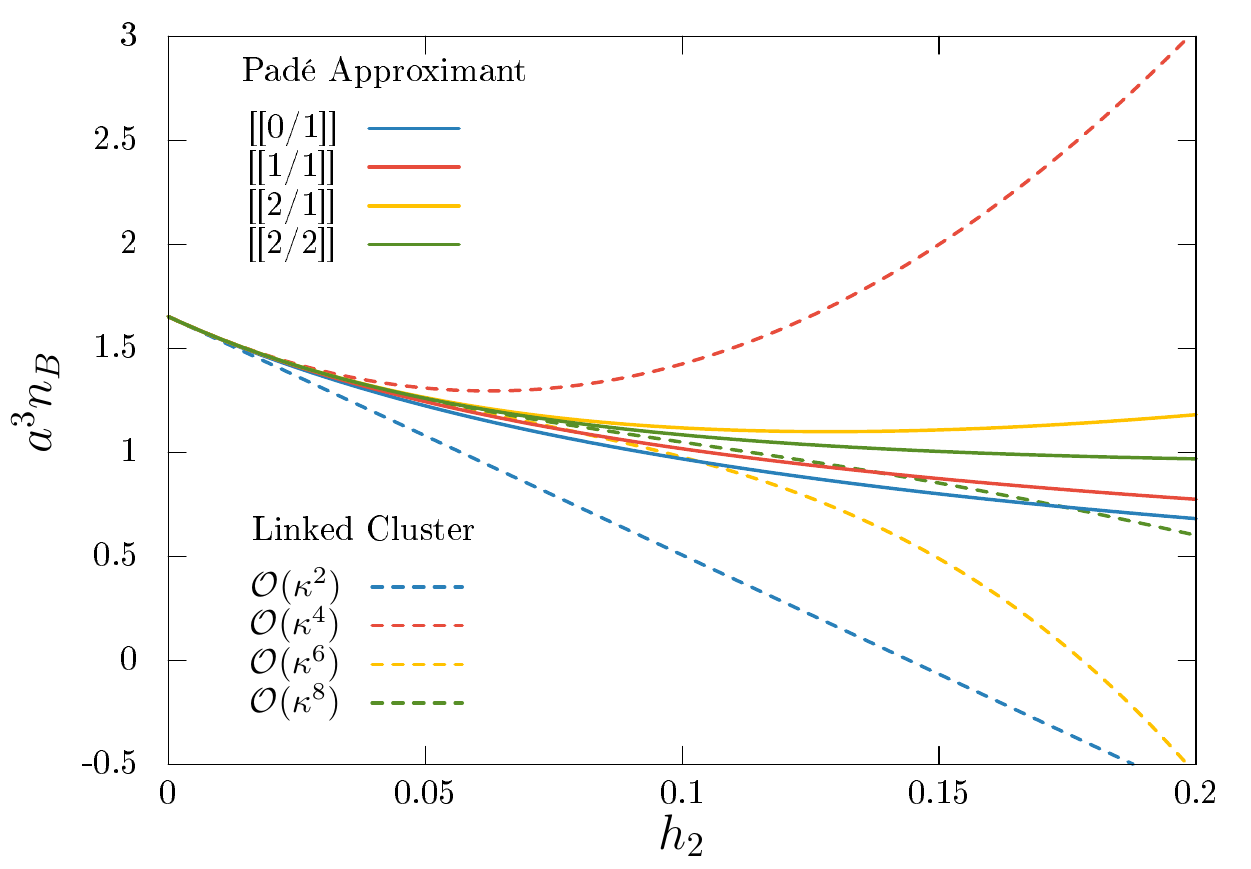}
\includegraphics[width=0.5\textwidth]{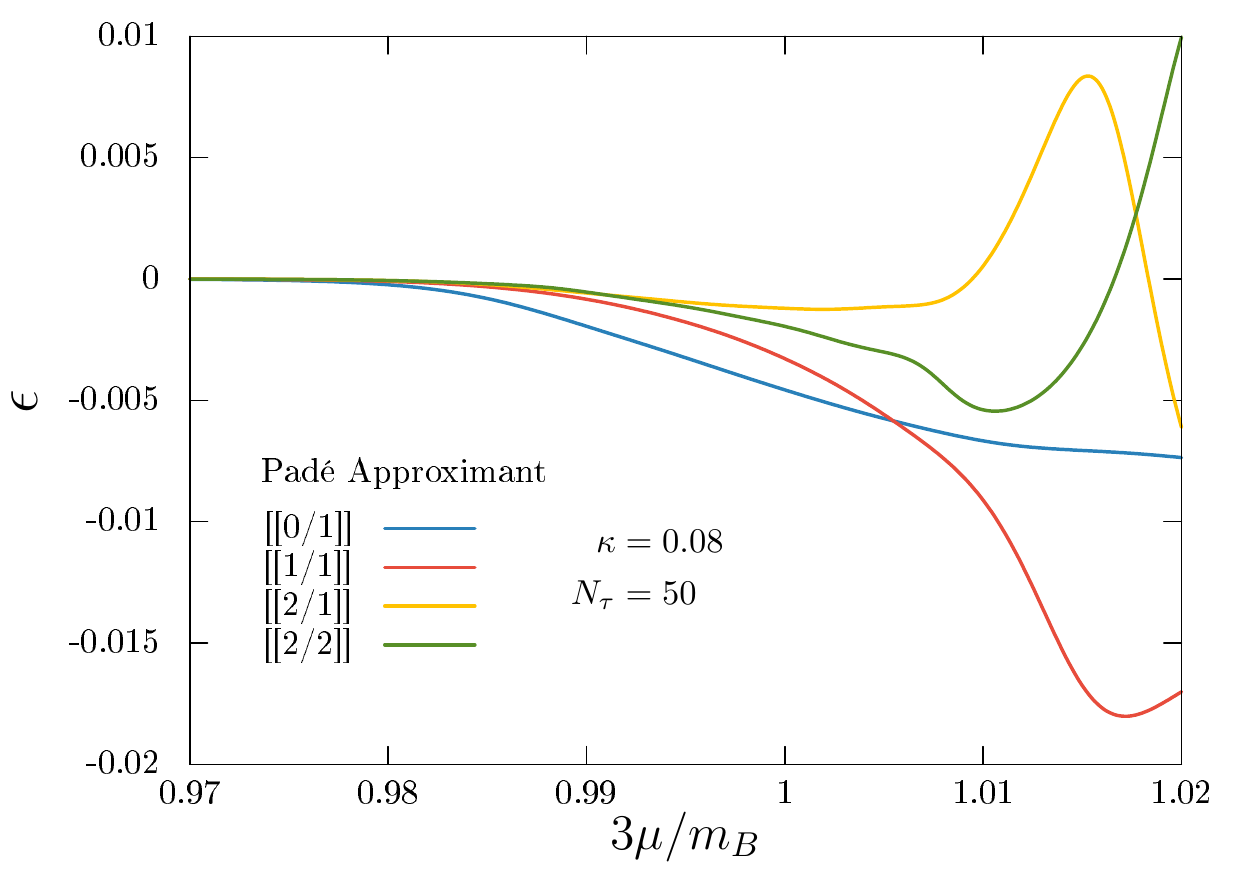}
}
\caption[]{Left: linked cluster expansion vs.~Pad\'e approximants up to the given order, 
$\beta=0, h_1=0.8$.
Right: Binding energy per nucleon.}
\label{fig:pade}
\end{figure}

Having coefficients order by order in $\kappa^2$, we next attempt to improve the convergence of
the series by constructing Pad\'e approximants. These have been used previously to
effectively resum the strong coupling and hopping expansion of the QCD deconfinement 
transition \cite{lp_deconf} and considerably improve convergence compared to the 
straight expansion \cite{scthermo}. Pad\'e approximants are rational functions constructed from power series 
of order $N$ and are defined as
\beq
[L,M](\kappa^2)=\frac{a_0+a_1\kappa^2 +\ldots + a_L(\kappa^2)^L}
{b_0+b_1\kappa^2 +\ldots + b_M(\kappa^2)^M}\;.
\eeq
In order to uniquely determine the coefficients $a_i, b_i$, it is necessary to have $L+M \leq N$, if $N$
 represents the highest available order of the original expansion. In this way an $[L,M]$ approximant
is correct up to, but not including ${\cal{O}}((\kappa^2)^{L+M+1})$, and larger 
approximants represent more expansion coefficients than smaller ones. Because of the choice
of $L,M$ for longer series, there is no unique Pad\'e approximant for a given series. We discard 
approximants that produce unphysical singularities through a vanishing denominator. Of the remaining
ones, the diagonal or close to diagonal ones are expected to be most reliable \cite{pade}.
 \fig\ref{fig:pade} (left) shows a considerably improved convergence 
 when using Pad\'e approximants for consecutive orders of the linked cluster expansion.
 
An important quantity characterising nuclear matter is the binding energy per nucleon. It
can be defined thermodynamically by the energy density minus mass density in the 
zero temperature limit,
 \beq
\epsilon(\mu)\equiv\lim_{T\rightarrow 0}\frac{e(T,\mu)-n_B(T,\mu)m_B}{n_B(T,\mu)m_B}
=\lim_{T\rightarrow 0}\frac{e(T,\mu)}{n_B(T,\mu)m_B}-1\;.
\label{eq:bind}
\eeq
In previous work we have shown numerically as well as analytically
to leading order in the hopping expansion, that this quantity displays the silver blaze property, 
i.e.~it is zero until the onset transition, where it becomes negative \cite{densek4}.
We can now extend this study to slightly larger densities. \fig\ref{fig:pade} shows the binding energy
extracted from the Pad\'e approximants to the partition function at various orders. While quantitative 
convergence breaks down shortly after the onset transition near $3\mu\sim m_B$, we obtain a new
qualitative result: in higher orders we see the binding energy becoming positive again with 
growing chemical potential, as is expected from nuclear physics. A minimum characterising
bulk nuclear density appears, which however is not yet settled quantitatively at the available orders.

\begin{figure}[t]
\centerline{
\includegraphics[width=0.5\textwidth]{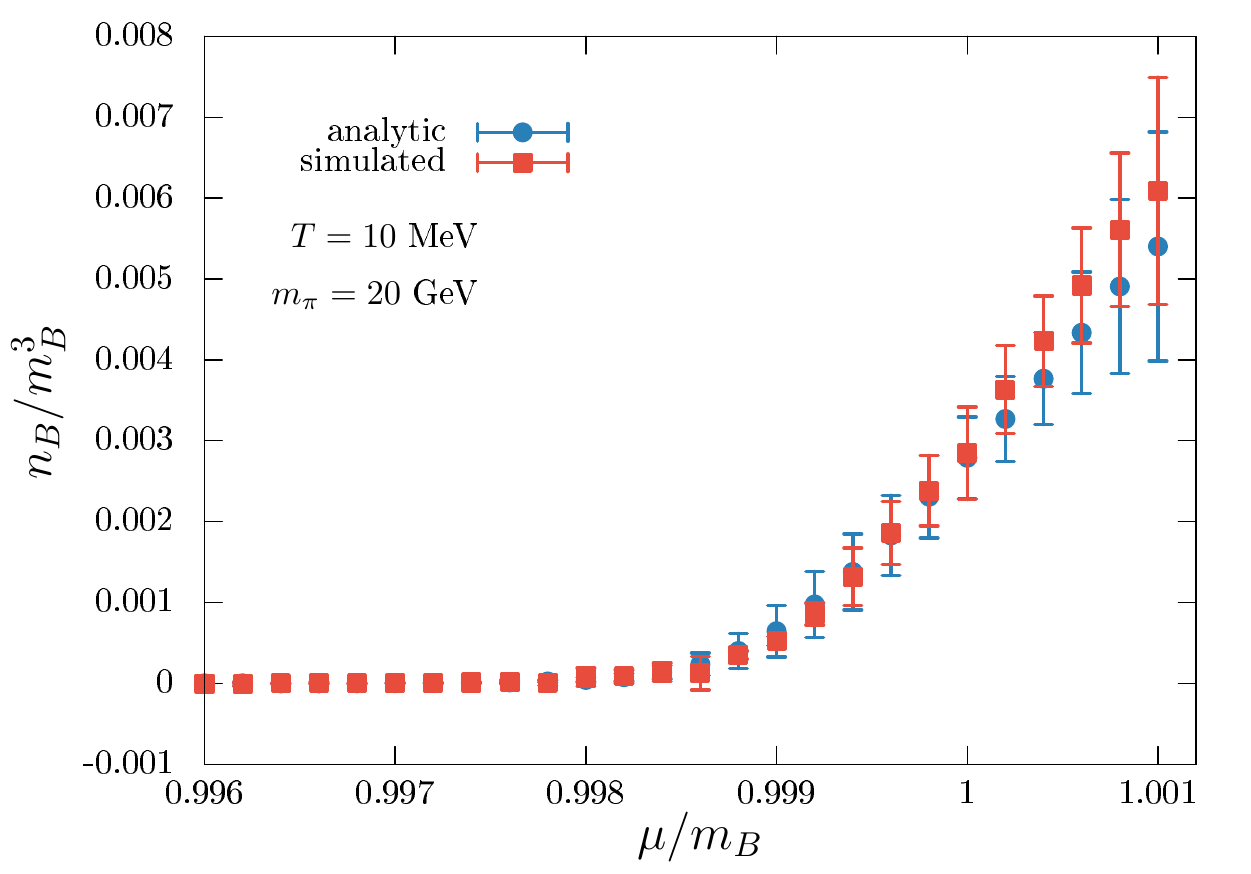}
}
\caption[]{Comparison of continuum extrapolations from numerical evaluation and linked
cluster expansion of the effective theory.}
\label{fig:press}
\end{figure}

Finally, we evaluate our effective theory by linked cluster expansion in a regime where we are able
to fully control it, which we again monitor by systematic errors taken as the difference between
consecutive orders. 
This is the heavy and cold regime which we have studied numerically in 
\cite{densek4}. 
We evaluate the equation of state for 6 different lattice spacings in the range of
$0.079\, \mathrm{fm} < a < 0.136\, \mathrm{fm}$ and perform a continuum extrapolation. 
The resulting baryon density as a function of chemical
potential
is shown in \fig\ref{fig:press} and quantitatively agrees with the numerical results.  

\section{A chain resummation} \label{sec:resum}

So far we have managed to reproduce most simulations of the effective theory with 
analytic calculations. In this section we will present a
resummation scheme for the analytic approach which extends 
beyond the reach of the numerical evaluation, in that it generates and then sums additional terms in 
the effective action.

\subsection{General idea} \label{sec:chain_intro}
We start with a motivating example. Consider the following four terms in
the effective action \eq\eqref{eq:effective_action},
\begin{equation}
  h_2 N_f \sum_{\mathrm{dof}} \; \begin{midtikzpicture}
    \node[circ] (n1) {1};
    \node[circ] (n2) at ([shift={(270:.75)}] n1) {1}
      edge[thick] (n1);
  \end{midtikzpicture}, \;
  - h_2^2 N_f\sum_{\mathrm{dof}}  \: \begin{midtikzpicture}
    \node[circ] (n1) {1};
    \node[circ] (n2) at ([shift={(60:{sqrt(2/3)})}] n1) {1}
      edge[thick] (n1);
    \node[circ] (n3) at ([shift={(300:{sqrt(2/3)})}] n2) {1}
      edge[thick] (n2);
  \end{midtikzpicture}, \;
  h_2^3 N_f\sum_{\mathrm{dof}} \: \begin{midtikzpicture}
    \node[circ] (n1) {1};
    \node[circ] (n2) at ([shift={(60:{sqrt(2/3)})}] n1) {1}
      edge[thick] (n1);
    \node[circ] (n3) at ([shift={(300:{sqrt(2/3)})}] n2) {1}
      edge[thick] (n2);
    \node[circ] (n4) at ([shift={(60:{sqrt(2/3)})}] n3) {1}
      edge[thick] (n3);
  \end{midtikzpicture}, \;
  - h_2^4 N_f\sum_{\mathrm{dof}} \: \begin{midtikzpicture}
    \node[circ] (n1) {1};
    \node[circ] (n2) at ([shift={(60:{sqrt(2/3)})}] n1) {1}
      edge[thick] (n1);
    \node[circ] (n3) at ([shift={(300:{sqrt(2/3)})}] n2) {1}
      edge[thick] (n2);
    \node[circ] (n4) at ([shift={(60:{sqrt(2/3)})}] n3) {1}
      edge[thick] (n3);
    \node[circ] (n5) at ([shift={(300:{sqrt(2/3)})}] n4) {1}
      edge[thick] (n4);
  \end{midtikzpicture} .
\end{equation}
It is easy to see that these four terms follow a common pattern to generate a chain. 
Each term extends the length of the chain by one node while
maintaining a common prefactor. Looking at the equations we see that every link in the chain adds a factor $h_2 W_{2,1}$ to the
term, along with the necessary spatial geometry. One can check with the terms up to order $\kappa^8$ in
\eq\eqref{eq:effective_action} that this holds not only for the simple one-string-chain shown above, 
but for all terms in the action with a
singly connected node, meaning nodes that correspond to a factor of $W_{1,1}$. The "chain" resummation can schematically be
represented as
\begin{equation} \label{eq:chain_schematic}
  \begin{tikzpicture}[baseline={([yshift=-.5ex]current bounding box.center)}]
    \node (l1) {$\mathcal{C}_0 = $};
    \node[circle,inner sep=2pt,minimum size=.5cm,draw,thick,right=5pt of l1] (g1) [align=center,scale=0.75] {rest of\\the term};
    \node[circ] at ([shift={(60:1.2)}] g1) {1}
      edge[thick] (g1);
    \node[circ] at ([shift={(300:1.2)}] g1) {1}
      edge[thick] (g1);
    \draw[->,thick] (2.5,0) -- +(2.1,0) node[midway,yshift=1ex,scale=0.75,font=\bfseries] {resummation};
    \begin{scope}[xshift=5.75cm]
      \node (l2) {$\mathcal{C}_n = $};
      \node[circle,inner sep=2pt,minimum size=.5cm,draw,thick,right=5pt of l2] (g2) [align=center,scale=0.75] {rest of\\the term};
      \node[circ] at ([shift={(60:1.2)}] g2) (g21) {1}
        edge[thick] (g2);
      \node[circ] at ([shift={(300:.75)}] g21) (g22) {1}
        edge[thick] (g21);
      \node[circle,scale=0.75] at ([shift={(60:.75)}] g22) (g23) {\dots}
        edge[thick] (g22);
      \node[circ] at ([shift={(300:.75)}] g23) (g24) {1}
        edge[thick] (g23);
      \node[circ] at ([shift={(60:.75)}] g24) (g25) {1}
        edge[thick] (g24);
      \node[circ] at ([shift={(300:1.2)}] g2) (g26) {1}
        edge[thick] (g2);
      \node[circ] at ([shift={(60:.75)}] g26) (g27) {1}
        edge[thick] (g26);
      \node[circle,scale=0.75] at ([shift={(300:.75)}] g27) (g28) {\dots}
        edge[thick] (g27);
      \node[circ] at ([shift={(60:.75)}] g28) (g29) {1}
        edge[thick] (g28);
      \node[circ] at ([shift={(300:.75)}] g29) (g210) {1}
        edge[thick] (g29);
    \end{scope}
  \end{tikzpicture}\;,
\end{equation}
where $n$ is the total number of links attached in the chains, and in the end we sum over all $n$. 
In the formulae this amounts to the substitution
\begin{equation} \label{eq:chain_subst}
  W_{1,1}(x) \to W_{1,1}(x) \sum_{n=0}^{\infty} \mathcal{G}(\{x_n\}) \prod_{i=1}^{n} (-h_2) W_{2,1}(x_i),
\end{equation}
where $\mathcal{G}(\{x_n\})$ contains the geometry of the chain.  Although the pattern is simple, 
to show that the prefactors
come out in a way that is summable is quite involved. 
A more thorough analysis of the resummation is given in appendix~\ref{app:chain}.

We now have a resummation of a class of diagrams to all orders in $\kappa$ 
for every order in the effective theory, but evaluating the final gauge
integral is impossible. This is because we need to sum over all geometries for all the terms in the resummation, which
cannot be evaluated analytically. To proceed, an additional constraint must be introduced, namely that only embeddings with the
same basic geometry as the starting structure are included.  
This implies that all nodes of the chain will be at separate lattice points,
and an $n+1$ long chain will result in the following integral (cf.~appendix~\ref{app:chain}),
\begin{align}
  \bigg( (2d) h_2 \int \mathrm{d} W \; \det\big[Q_{\mathrm{stat}}\big]^{2 N_f} W_{2,1} \bigg)^n 
  (2d) h_2 \int \mathrm{d} W\; \det&\big[Q_{\mathrm{stat}}\big]^{2 N_f} W_{1,1}  \nonumber\\
  &\equiv (2d h_2)^{n+1} I_{2,1}^n  I_{1,1}.
\end{align}
The $I_{n,m}$ denote the integrals over $W$ on a lattice point occupied by a single $W_{n,m}$ factor.
To carry out the integral one needs again to convert from $W$ to $L$ using the generating
functional presented in appendix~\ref{app:genFunc}.
The full chain will then give
\begin{equation} \label{eq:chain_resummation_formula}
  (2d) h_2 I_{1,1} \sum_{n=0}^{\infty} (-(2d) h_2)^n I_{2,1}^n = \frac{(2d) h_2 I_{1,1}}{1 + (2d)h_2 I_{2,1}}\:.
\end{equation}

\subsection{Validity of the combinatorics}

\begin{figure}
  {\centering
    \includegraphics[width=.6\textwidth]{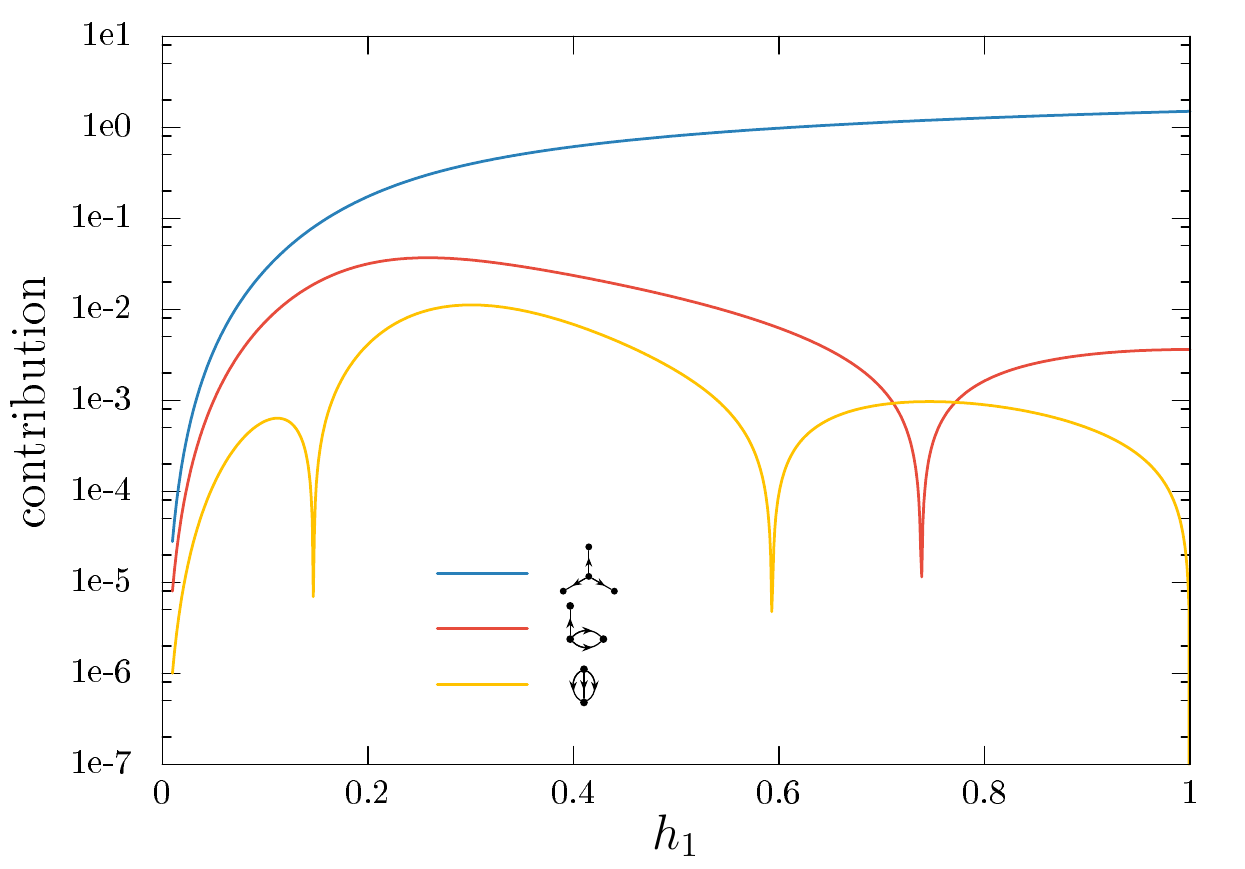}
    \par}
  \caption[]{Normalised contributions to the free energy from the embeddings of the
    \scalebox{0.75}{\begin{midtikzpicture}
    \node[circ] (n1) {1};
    \node[circ] at ([shift={(-30:.5)}] n1) {1}
      edge[thick] (n1);
    \node[circ] at ([shift={(90:.5)}] n1) {1}
      edge[thick] (n1);
    \node[circ] at ([shift={(210:.5)}] n1) {1}
      edge[thick] (n1);
  \end{midtikzpicture}} term}
  \label{fig:overlapping}
\end{figure}

Carrying out the chain resummation and embedding it on the simplest cluster expansion graph 
as outlined above does introduce small systematic
errors. This is because the embedding factor of $(2d)$ also counts graphs belonging to self-overlapping
chains and thus is too large. 
However, the difference of the overcounted contributions and the actual
self-overlapping embeddings, which represents the error, results in cumulants of the participating factors. In
\fig\ref{fig:overlapping} examples of these cumulants from the overlapping embeddings of a 
specific term have been plotted. We see that the non-overlapping 
configuration is orders of magnitude larger than the overlapping ones, especially as the
fugacity $h_1$ approaches $1$. The same holds for the other types of graphs as well.
This behaviour is due to the fact that the integrals
\begin{equation*}
  \int \mathrm{d} L \, W_{1,1}^n \hspace{1cm}\text{and}\hspace{1cm} \Big( \int \mathrm{d} L \, W_{1,1} \Big)^n
\end{equation*}
are of the same magnitude, resulting in cancellations in the cumulants. Therefore, the error introduced by
our embedding is small.

\subsection{Results}

\begin{figure}[t]
  {\centering
      \includegraphics[width=.5\textwidth]{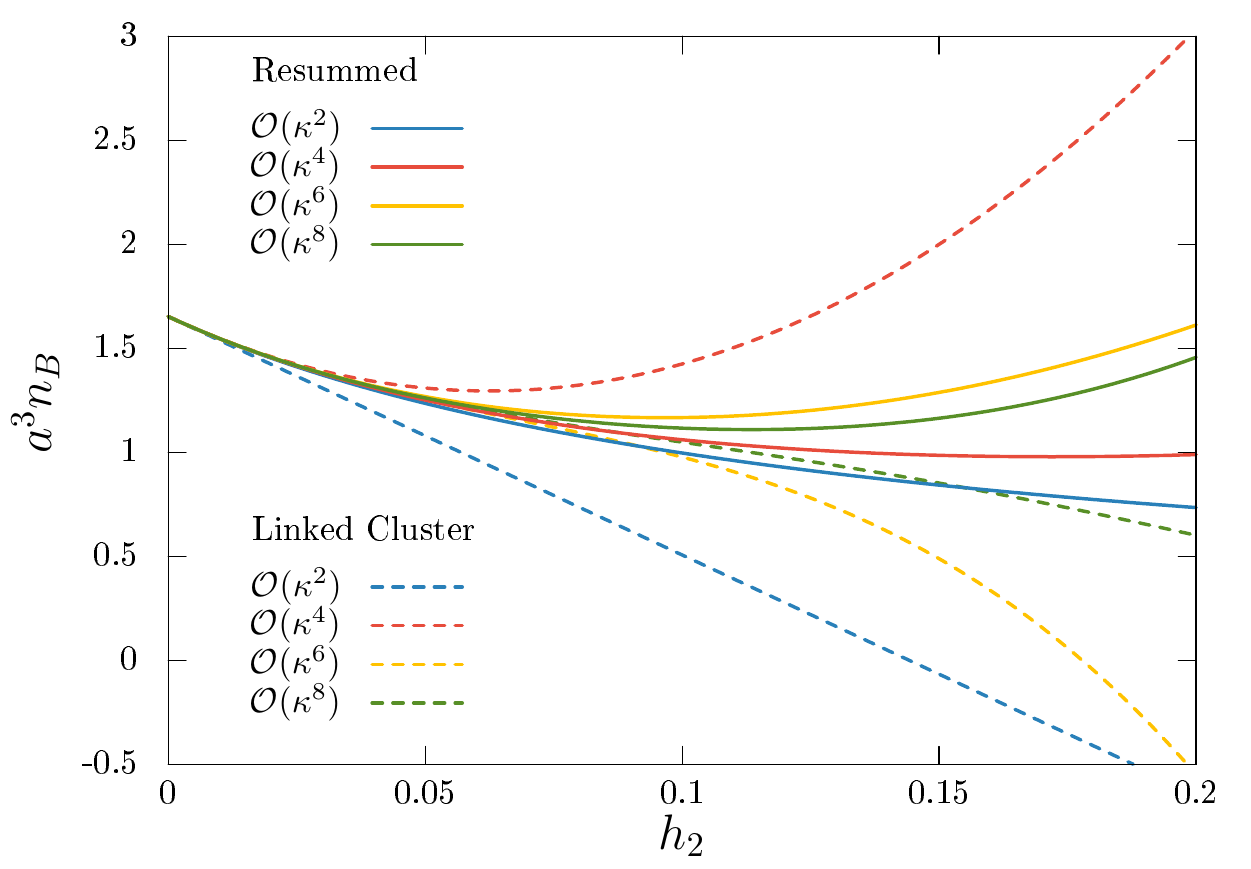}
      \includegraphics[width=.5\textwidth]{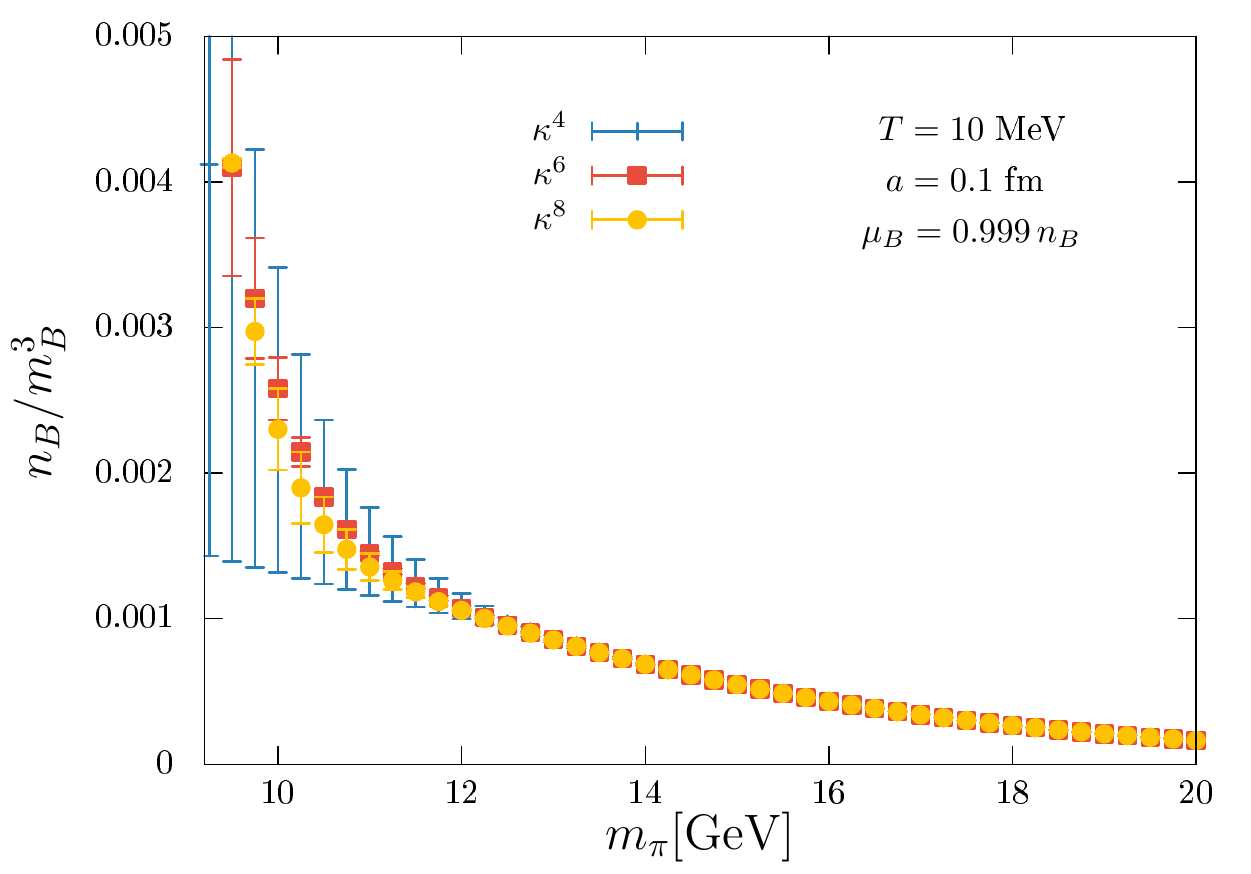}
  }
  \caption[]{Left: Convergence of the resummation-improved results. 
   Right: Baryon number density as a function of pion mass including chain resummation. The convergence region 
   is extended compared to \fig\ref{fig:mass}.}
  \label{fig:improved_convergence}
\end{figure}

With a new and extended effective action we can redo the calculations from 
section~\ref{sec:cle_results}. In
\fig\ref{fig:improved_convergence} (left) the increase in convergence due to the resummation scheme is 
clearly visible, more than
doubling the convergence region in $h_2$. 
Matching the plot with that of \fig\ref{fig:pade}, one sees a comparable increase in
convergence to that from the Pad\'{e} approximation. This is both expected and reassuring as 
both approaches produce rational
expressions, and the superior convergence of the Pad\'{e} is expected due to the fact that it is not 
restricted to a particular
class of diagrams and might therefore predict higher order behaviour. 

In
\fig\ref{fig:improved_convergence} (right) we have repeated the pion mass 
convergence plot and one can see that the resummation extends the
convergence region in a natural way.  
\begin{figure}[t]
\centerline{
      \includegraphics[width=.5\textwidth]{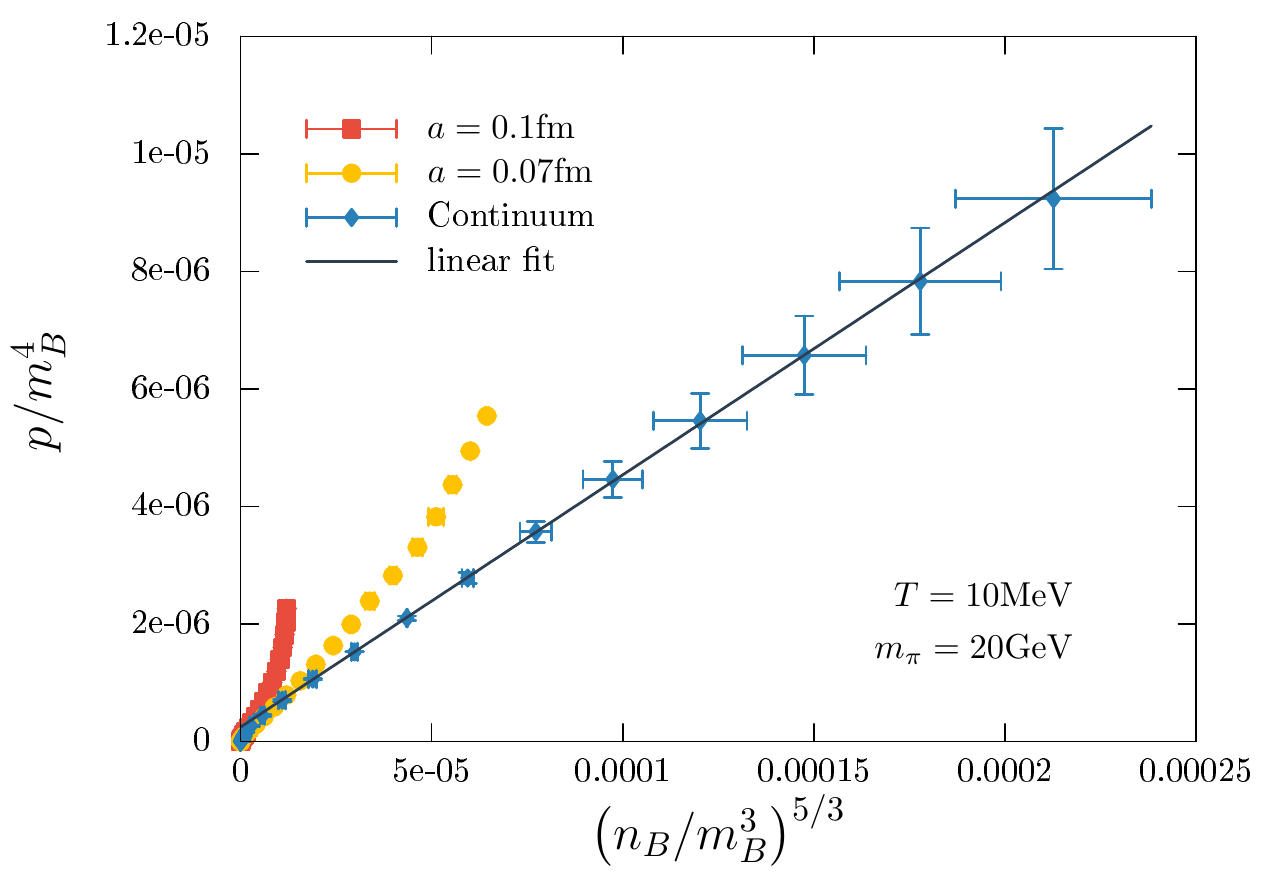}
  }
  \caption[]{%
    Pressure as a function of baryon number density to the power $5/3$, analytically calculated with the chain resummation. Error
    bars reflect different orders in the expansion as well as varying number of points used in the continuum extrapolation.}
  \label{fig:eos}
\end{figure}

We now give our final result, the equation of state for nuclear matter with heavy quarks calculated fully 
analytically, \fig\ref{fig:eos}. The error bars represent the uncertainty resulting from continuum extrapolations
including a varying number of points. The line represents a fit to a polytropic equation of state
for non-relativistic fermions,
\beq
\frac{p}{m_B^4}\sim  0.0429(29) \left (\frac{n_B}{m_B^3}\right)^{5/3}\;.
\eeq
This might of course be expected on physical grounds: the bosonic baryons condense and do not 
contribute to the pressure, which is thus due to the heavy fermionic baryons. However, computationally this 
is a very remarkable result. Firstly, we have started with an action in terms of quarks and gluons 
and expanded about the strong coupling limit. The emergence of baryonic degrees of freedom 
with a weak attractive interaction \cite{densek4}
is completely dynamical and a result of the calculation. Secondly, the 
equation of state for {\it any} finite lattice spacing shows saturation and thus lattice fermions 
do {\it not} feature a polytropic equation of state, which is also clearly visible in \fig\ref{fig:eos}. 
The fact that our continuum extrapolation is well described
by a physically sensible polytrope then appears to be an endorsement of our
calculation. 
It will now be very interesting to
investigate the prefactor of the polytropic behaviour, 
which must depend on the contributing degrees of freedom, 
their masses and interactions.  

\section{Conclusions}

We have extended a previously derived three-dimensional effective lattice action for QCD 
to the order $u^5\kappa^8$ in the cold limit with a combined character and hopping parameter expansion, starting from the full
Wilson action.  The effective action has a sign problem mild enough to permit controlled
simulations of the cold and dense regime for heavy quarks, where our expansion is valid. The additional
orders fully confirm our previous results and demonstrate that the systematics in this approach can be 
monitored and controlled. 

In the second part of the paper we have exploited the fact that our effective theory formally corresponds to 
a spin model with multi-point couplings over various ranges, which we were able to map onto a linked cluster 
expansion. This permits a completely analytic evaluation of cold and dense thermodynamics 
which is entirely unaffected by the sign problem. The convergence of the linked cluster
expansion is excellent and fully reproduces the numerical simulations of the effective theory in the range of
its validity. In this framework we were furthermore able to identify a class of diagrams consisting of chains
of arbitrary length representing meson exchange, which can be summed up to all orders in the hopping
expansion. 

For sufficiently heavy quarks, continuum extrapolations of the thermodynamical
functions are possible to provide the equation of state for heavy bulk nuclear matter. Our final result
is consistent with a polytropic system of non-relativistic fermions as expected on physical grounds.

\section*{Acknowledgements}
We thank Wolfgang Unger for numerous enlightening discussions, and Eduardo Fraga 
for alerting us to polytropic equations of state.
J.G.~and O.P.~were supported by the German BMBF, No. 06FY7100 as well as
Helmholtz International Center for FAIR within the LOEWE program of the State of Hesse.
The authors acknowledge computing time granted on the FUCHS cluster of 
Goethe-University Frankfurt. 

\newpage
\begin{appendices}

\section{Effective action in the cold limit} \label{app:action}
The effective action to order $\kappa^8$ and to leading order in $\frac{1}{N_{\tau}}$ reads
\begin{equation}
  S_{\mathrm{eff}} = S_2 + S_4 + S_6 + S_8 + \mathcal{O}\big(\kappa^{10}, \frac{1}{N_{\tau}}\big)
\end{equation}

$S_2 = $ \vspace*{-10pt}

\begin{equation}
   + h_2 N_f \sum_x \sum_i W_{1,1}(x) W_{1,1}(x+i) \label{eq:expansionBegin}
\end{equation}

$S_4 = $ \vspace*{-10pt}

\begin{subequations}
\begin{align}
  &-h_2^2 N_f \sum_x \sum_{i,j} W_{2,1}(x) W_{1,1}(x+i) W_{1,1}(x+j) \\
  &-h_2^2 N_f^2 \sum_x \sum_{i} W_{2,1}(x) W_{2,1}(x+i)
\end{align}
\end{subequations}

$S_6 = $ \vspace*{-10pt}

\begin{subequations}
\begin{align}
  +\frac{1}{3}h_2^3 N_f &\sum_x \sum_{i,j,k} \big[ W_{3,1}(x) - W_{3,2}(x) \big] W_{1,1}(x+i) \label{eq:w3w1w1w1}
   W_{1,1}(x+j) W_{1,1}(x+k) \\
  +h_2^3 N_f &\sum_x \sum_{i,j,k} W_{2,1}(x) W_{2,1}(x+i) W_{1,1}(x+i+j) W_{1,1}(x+k) \\
  +2h_2^3 N_f^2 &\sum_x \sum_{i,j} \big[ W_{3,1}(x) - W_{3,2}(x) \big] W_{2,1}(x+i) W_{1,1}(x+j) \\
  +\frac{1}{6}h_2^3 N_f &\sum_x \sum_i \big[ W_{3,1}(x) W_{3,1}(x+i) + W_{3,2}(x) W_{3,2}(x+i) \big] \\
  -\frac{4}{3}h_2^3 N_f^3 &\sum_x \sum_i W_{3,1}(x) W_{3,2}(x+i) \label{eq:expansionEnd}
\end{align}
\end{subequations}

$S_8 = $ \vspace*{-10pt}

{ \allowdisplaybreaks
\begin{subequations}
\begin{align}
  +\frac{1}{12} h_2^4 N_f &\smash[b]{\sum_x} \smash[b]{\sum_{i,j,k,l}}
    \big[ W_{4,1}(x) - 4 W_{4,2}(x) + W_{4,3}(x) \big] W_{1,1}(x+i) W_{1,1}(x+j) \nonumber\\
  & \hspace{4em} \times W_{1,1}(x+k) W_{1,1}(x+l) \\
  +h_2^4 N_f &\smash[b]{\sum_x} \smash[b]{\sum_{i,j,k,l}}
    \big[ W_{3,1}(x) - W_{3,2}(x)  \big] W_{2,1}(x+i) W_{1,1}(x+i+j) \nonumber\\
  & \hspace{4em} \times W_{1,1}(x+k) W_{1,1}(x+l) \\
  +h_2^4 N_f &\smash[b]{\sum_x} \smash[b]{\sum_{i,j,k,l}} W_{2,1}(x) W_{2,1}(x+i) W_{2,1}(x+j) \nonumber\\
  & \hspace{4em} \times W_{1,1}(x+i+k) W_{1,1}(x+j+l) \\
  +h_2^4 N_f^2 &\smash[b]{\sum_x} \smash[b]{\sum_{i,j,k}} \big[ W_{4,1}(x) - 4 W_{4,2}(x) + W_{4,3}(x) \big] W_{2,1}(x+i) \nonumber\\
  & \hspace{4em} \times W_{1,1}(x+j) W_{1,1}(x+k) \\
  +h_2^4 N_f^2 &\smash[b]{\sum_x} \smash[b]{\sum_{i,j,k}}
    \big[ W_{3,1}(x) W_{3,1}(x+i) - 2 W_{3,1}(x) W_{3,2}(x+i) + W_{3,2}(x) W_{3,2}(x+i)\big] \nonumber\\
  & \hspace{4em} \times W_{1,1}(x+j) W_{1,1}(x+i+k) \\
  +2 h_2^4 N_f^2 &\sum_x \sum_{i,j,k}
    \big[ W_{3,1}(x) - W_{3,2}(x) \big] W_{2,1}(x+i) W_{2,1}(x+j) W_{1,1}(x+j+k) \\
  + h_2^4 N_f^2 &\sum_x \sum_{i,j} W_{2,1}(x) W_{2,1}(x) W_{2,1}(x+i) W_{2,1}(x+j) \\
  + \frac{1}{2} h_2^4 N_f^2 &\sum_x \sum_{i,j} W_{2,1}(x) W_{2,1}(x+i) W_{2,1}(x+j) W_{2,1}(x+i+j) \\
  +\frac{1}{3} h_2^4 N_f &\smash[b]{\sum_x} \smash[b]{\sum_{i,j}}
  \big[ W_{4,1}(x) W_{3,1}(x+i) - 2 W_{4,2}(x) W_{3,1}(x+i) + 2 W_{4,2}(x) W_{3,2}(x+i) \nonumber\\
  & \hspace{4em} - W_{4,3}(x) W_{3,2}(x+i) \big] W_{1,1}(x+j) \\
  -\frac{4}{3} h_2^4 N_f^3 &\smash[b]{\sum_x} \smash[b]{\sum_{i,j}}
    \big[ 2 W_{4,2}(x) W_{3,1}(x+i) - W_{4,3}(x) W_{3,1}(x+i) + W_{4,1}(x) W_{3,2}(x+i) \nonumber\\
  & \hspace{4em} - 2 W_{4,2}(x) W_{3,2}(x+i) \big] W_{1,1}(x+j) \\
  + \frac{1}{12} h_2^4 N_f &\sum_x \sum_{i,j}
    \big[ W_{4,1}(x) - 4 W_{4,2}(x) + W_{4,3}(x) \big] W_{2,1}(x+i) W_{2,1}(x+j) \\
  + \frac{2}{3} h_2^4 N_f^3 &\sum_x \sum_{i,j}
    \big[ W_{4,1}(x) - 4 W_{4,2}(x) + W_{4,3}(x) \big] W_{2,1}(x+i) W_{2,1}(x+j) \\
  + \frac{1}{12} h_2^4 N_f^2 &\smash[b]{\sum_x} \smash[b]{\sum_i}
    \big[ W_{4,1}(x) W_{4,1}(x+i) + 12 W_{4,2}(x) W_{4,2}(x+i) \nonumber\\
  & \hspace{4em} + W_{4,3}(x) W_{4,3}(x+i) \big] \\
  + \frac{2}{3} h_2^4 N_f^4 &\sum_x \sum_i \big[ W_{4,1}(x) W_{4,3}(x+i) + 2 W_{4,2}(x) W_{4,2}(x+i) \big] \\
  - \frac{2}{3} h_2^4 N_f^2 &\sum_x \sum_i \big[ W_{4,1}(x) W_{4,2}(x+i) + W_{4,2}(x) W_{4,3}(x+i) \big]
\end{align}
\end{subequations}
}
where the sums of $\{i,j,k,l\}$ go over all spatial directions. At this order in $\frac{1}{N_{\tau}}$ all gauge corrections come
from the rescaling of the coupling constants $h_1(u,\kappa)$ and $h_2(u,\kappa)$. Moving away from this limit will result in gauge corrections that
depend on the spatial geometry of the various terms.

\section{\texorpdfstring{Generating functional for the $W_{n,m}$ terms}{Generating functional for the Wnm terms}} \label{app:genFunc}

In our calculation of the effective theory we need to convert terms on the form:
\begin{equation}
  \tr \frac{\big(h_1 W\big)^n}{\big(1 + h_1 W\big)^m}
\end{equation}
to be functions of the Polyakov loop $L = \tr W$. We can accomplish this using the following generating functional,
\begin{equation}
  G[\alpha,\beta] = \tr \ln \big( \alpha + \beta h_1 W \big)\;,
\end{equation}
together with the trace-log identity and the expression
\begin{equation}
  \det \big( \alpha + \beta h_1 W \big) = \alpha^3 + \alpha^2 \beta h_1 L + \alpha \beta^2 h_1^2 L^* 
  + \beta^3 h_1^3\;,
\end{equation}
which is derived from the Calay-Hamilton theorem. It is then straightforward to show that
\begin{equation}
  \frac{(-1)^{n+m-1}}{(n+m-1)!} \bigg( \frac{\partial}{\partial \alpha} \bigg)^n 
  \bigg( \frac{\partial}{\partial \beta} \bigg)^m
  G[\alpha,\beta] \Bigg|_{\alpha=\beta\mathrlap{=1}} =
   \tr \frac{\big(h_1 W\big)^m}{\big(1 + h_1 W\big)^{n+m}}\;.
\end{equation}

\section{The chain resummation \label{app:chain}}

\subsection{Higher order terms in the effective theory}

As a prerequisite to our chain resummation, we need to understand the systematics and coefficients of 
the terms in the effective action that we wish to sum up. 
To this end, we recall the expansion of the kinetic
determinant, \eqref{eq_detqkin}, 
\begin{align}
\det[Q_\mathrm{kin}]&=  \exp \Bigg\{ -  \sum_{n=1}^{\infty} \frac{1}{n} \tr \big(P + M\big)^n \Bigg\} \nonumber \\
   &= 1 -  \tr PM + \frac{1}{2!} \Big( \tr PM \Big)^2 -  \tr PPMM - \frac{1}{2} 
     \tr PMPM + \mathcal{O}\big(\kappa^6\big)\;,
  \label{eq:eff_action_expansion}
\end{align}
in terms of forward and backward spatial hopping matrices \cite{densek4}
\begin{align}
  P_{ab}^{\alpha\beta}(x,y) &= 
    \kappa \, T_{ac}^{\alpha\delta}(x,z) \sum_{\mathclap{i \in \{\hat{x},\hat{y},\hat{z}\}}} \big(1 - \gamma_i \big)^{\delta\beta}
    U_{i,cb}(z) \delta_{\tau_z, \tau_y} \delta_{\vec{z}+i,\vec{y}}\;,\\
  M_{ab}^{\alpha\beta}(x,y) &= 
    \kappa \, T_{ac}^{\alpha\delta}(x,z) \sum_{\mathclap{i \in \{\hat{x},\hat{y},\hat{z}\}}} \big(1 + \gamma_i \big)^{\delta\beta}
    U_{i,cb}^{\dagger}(z) \delta_{\tau_z,\tau_y} \delta_{\vec{z}-i,\vec{y}}\;.
\end{align}
Here $T$ is the static propagator whose Dirac and colour structure neatly separate,
\begin{equation} \label{eq:static_propagator}
  T_{ab}^{\alpha\beta}(x,y) = \delta_{x,y} \delta_{a,b} \delta_{\alpha,\beta} + \big(1 - \gamma_0\big)^{\alpha\beta} 
    B_{ab}(\tau_x,\tau_y) \delta_{\vec{x},\vec{y}}\;,
\end{equation}
and the expression for $B$ can be found in \cite{densek4}. 
After the integration over the
spatial gauge links, only a few combinations give non-zero contributions to the final result. 
This is because gauge integrals over many 
combinations of link variables vanish \cite{Creutz:1978ub},
\begin{equation}
  \int \mathrm{d} U \, U^n \big(U^{\dagger}\big)^m = 0, \; \text{  if  } \; n + 2m \neq 0 \pmod{3}.
\end{equation}
Since $P \sim U$ and $M \sim U^{\dagger}$, we can identify
the non-vanishing terms as those, where at every spatial gauge link 
the number of $P$'s and $M$'s satisfy
\begin{equation}
  (\mbox{number of}\; P's)+2\times (\mbox{number of}\; M's)=0 \pmod{3} .
\end{equation}
We define contractions to be those terms for which the 
gauge integration gives non-vanishing contributions. 
Thus a contraction identifies the matrix products with such spatial and temporal coordinates that the
gauge links overlap, e.g., 
\begin{align}
  &\begin{tikzpicture}[
      baseline = .5ex,
      every node/.style = {anchor=south, text depth=0.35ex, inner sep=1pt}
    ]
    \node (one) {$P(x,y)$};
    \node (two) [right=0pt of one]{$M(z,w)$};
    \path ([xshift=7pt]one.north west) edge [skip loop=6pt] ([xshift=7pt]two.north west);
  \end{tikzpicture} = P(x,y) M(z,w) \,\delta_{\vec{y}\vec{z}} \,\delta_{\tau_y\tau_w} \,.
  \label{eq:contraction_def}
\end{align}
Contractions can consist of any number of matrices/links. As an example, all contractions of the
$\mathcal{O}\big(\kappa^6\big)$ term $\tr PMPMPM$ are given in table~\ref{tab:contractions}. 
Since one contraction fixes the space
and time degrees of freedom of all participating matrices, a higher  
number of separate contractions per term has more
degrees of freedom left for the remaining traces, and in particular more terms in the temporal
direction. Summing over the latter, we  can categorise the terms in table~\ref{tab:contractions} as 
in the following examples:
\begin{subequations}
  \tikzset{
    every node/.style = {anchor=south, inner sep=1pt}
  }
  \begin{align}
    \sum_{\mathclap{\tau_1,\tau_2,\tau_3}} \begin{tikzpicture}[baseline=.5ex]
      \node (a) {$P$};
      \node (b) [right=0pt of a]{$M$};
      \node (c) [right=0pt of b]{$P$};
      \node (d) [right=0pt of c]{$M$};
      \node (e) [right=0pt of d]{$P$};
      \node (f) [right=0pt of e]{$M$};
      \path (a.north) edge [skip loop=6pt] (b.north);
      \path (c.north) edge [skip loop=6pt] (d.north);
      \path (e.north) edge [skip loop=6pt] (f.north);
    \end{tikzpicture} \propto N_{\tau}^3\;,\\
    \sum_{\tau_1,\tau_2}\begin{tikzpicture}[baseline=.5ex]
      \node (a) {$P$};
      \node (b) [right=0pt of a]{$M$};
      \node (c) [right=0pt of b]{$P$};
      \node (d) [right=0pt of c]{$M$};
      \node (e) [right=0pt of d]{$P$};
      \node (f) [right=0pt of e]{$M$};
      \path (a.north) edge [skip loop=6pt] (b.north);
      \path (c.north) edge [skip loop=6pt] (f.north);
      \path (d.north) edge [skip loop=6pt] (e.north);
    \end{tikzpicture} \propto N_{\tau}^2\;,\\
    \sum_{\tau_1,\tau_2}\begin{tikzpicture}[baseline=.5ex]
      \node (a) {$P$};
      \node (b) [right=0pt of a]{$M$};
      \node (c) [right=0pt of b]{$P$};
      \node (d) [right=0pt of c]{$M$};
      \node (e) [right=0pt of d]{$P$};
      \node (f) [right=0pt of e]{$M$};
      \path (a.north) edge [skip loop=4pt] (c.north);
      \path (c.north) edge [skip loop=4pt] (e.north);
      \path (b.north) edge [skip loop=8pt] (d.north);
      \path (d.north) edge [skip loop=8pt] (f.north);
    \end{tikzpicture} \propto N_{\tau}^2\;,\\
    \sum_{\tau_1}\begin{tikzpicture}[baseline=.5ex]
      \node (a) {$P$};
      \node (b) [right=0pt of a]{$M$};
      \node (c) [right=0pt of b]{$P$};
      \node (d) [right=0pt of c]{$M$};
      \node (e) [right=0pt of d]{$P$};
      \node (f) [right=0pt of e]{$M$};
      \path (a.north) edge [skip loop=6pt] (d.north);
      \path (b.north) edge [skip loop=6pt] (e.north);
      \path (c.north) edge [skip loop=6pt] (f.north);
    \end{tikzpicture} \propto N_{\tau}\;.
  \end{align}
\end{subequations}
In the low temperature limit, where $N_{\tau} \to \infty$, the contractions consisting of pairs 
will dominate the result and we restrict our attention to those for the remainder.
\begin{table}
\begin{center}
  \tikzset{
    every node/.style = {anchor=south, text depth=0.35ex, inner sep=1pt}
  }
  \begin{tabular}{cccc}
    \begin{tikzpicture}[baseline={(a.south)}]
      \node (a) {$P$};
      \node (b) [right=0pt of a]{$M$};
      \node (c) [right=0pt of b]{$P$};
      \node (d) [right=0pt of c]{$M$};
      \node (e) [right=0pt of d]{$P$};
      \node (f) [right=0pt of e]{$M$};
      \path (a.north) edge [skip loop=6pt] (b.north);
      \path (c.north) edge [skip loop=6pt] (d.north);
      \path (e.north) edge [skip loop=6pt] (f.north);
    \end{tikzpicture} &
    \begin{tikzpicture}[baseline={(a.south)}]
      \node (a) {$P$};
      \node (b) [right=0pt of a]{$M$};
      \node (c) [right=0pt of b]{$P$};
      \node (d) [right=0pt of c]{$M$};
      \node (e) [right=0pt of d]{$P$};
      \node (f) [right=0pt of e]{$M$};
      \path (a.north) edge [skip loop=8pt] (d.north);
      \path (c.north) edge [skip loop=4pt] (b.north);
      \path (e.north) edge [skip loop=4pt] (f.north);
    \end{tikzpicture} &
    \begin{tikzpicture}[baseline={(a.south)}]
      \node (a) {$P$};
      \node (b) [right=0pt of a]{$M$};
      \node (c) [right=0pt of b]{$P$};
      \node (d) [right=0pt of c]{$M$};
      \node (e) [right=0pt of d]{$P$};
      \node (f) [right=0pt of e]{$M$};
      \path (a.north) edge [skip loop=4pt] (b.north);
      \path (c.north) edge [skip loop=8pt] (f.north);
      \path (e.north) edge [skip loop=4pt] (d.north);
    \end{tikzpicture} &
    \begin{tikzpicture}[baseline={(a.south)}]
      \node (a) {$P$};
      \node (b) [right=0pt of a]{$M$};
      \node (c) [right=0pt of b]{$P$};
      \node (d) [right=0pt of c]{$M$};
      \node (e) [right=0pt of d]{$P$};
      \node (f) [right=0pt of e]{$M$};
      \path (a.north) edge [skip loop=12pt] (f.north);
      \path (c.north) edge [skip loop=4pt] (d.north);
      \path (e.north) edge [skip loop=8pt] (b.north);
    \end{tikzpicture} \\
    \begin{tikzpicture}[baseline={(a.south)}]
      \node (a) {$P$};
      \node (b) [right=0pt of a]{$M$};
      \node (c) [right=0pt of b]{$P$};
      \node (d) [right=0pt of c]{$M$};
      \node (e) [right=0pt of d]{$P$};
      \node (f) [right=0pt of e]{$M$};
      \path (a.north) edge [skip loop=8pt] (f.north);
      \path (c.north) edge [skip loop=4pt] (b.north);
      \path (e.north) edge [skip loop=4pt] (d.north);
    \end{tikzpicture} &
    \begin{tikzpicture}[baseline={(a.south)}]
      \node (a) {$P$};
      \node (b) [right=0pt of a]{$M$};
      \node (c) [right=0pt of b]{$P$};
      \node (d) [right=0pt of c]{$M$};
      \node (e) [right=0pt of d]{$P$};
      \node (f) [right=0pt of e]{$M$};
      \path (a.north) edge [skip loop=4pt] (d.north);
      \path (c.north) edge [skip loop=12pt] (f.north);
      \path (e.north) edge [skip loop=8pt] (b.north);
    \end{tikzpicture} &
    \begin{tikzpicture}[baseline={(a.south)}]
      \node (a) {$P$};
      \node (b) [right=0pt of a]{$M$};
      \node (c) [right=0pt of b]{$P$};
      \node (d) [right=0pt of c]{$M$};
      \node (e) [right=0pt of d]{$P$};
      \node (f) [right=0pt of e]{$M$};
      \path (a.north) edge [skip loop=4pt] (c.north);
      \path (c.north) edge [skip loop=4pt] (e.north);
      \path (b.north) edge [skip loop=8pt] (d.north);
      \path (d.north) edge [skip loop=8pt] (f.north);
    \end{tikzpicture} &
    \begin{tikzpicture}[baseline={(a.south)}]
      \node (a) {$P$};
      \node (b) [right=0pt of a]{$M$};
      \node (c) [right=0pt of b]{$P$};
      \node (d) [right=0pt of c]{$M$};
      \node (e) [right=0pt of d]{$P$};
      \node (f) [right=0pt of e]{$M$};
      \path (a.north) edge [skip loop=6pt] (d.north);
      \path (c.north) edge [skip loop=6pt] (b.north);
      \path (e.north) edge [skip loop=6pt] (f.north);
    \end{tikzpicture} \\
    \begin{tikzpicture}[baseline={(a.south)}]
      \node (a) {$P$};
      \node (b) [right=0pt of a]{$M$};
      \node (c) [right=0pt of b]{$P$};
      \node (d) [right=0pt of c]{$M$};
      \node (e) [right=0pt of d]{$P$};
      \node (f) [right=0pt of e]{$M$};
      \path (a.north) edge [skip loop=6pt] (b.north);
      \path (c.north) edge [skip loop=6pt] (f.north);
      \path (d.north) edge [skip loop=6pt] (e.north);
    \end{tikzpicture} &
    \begin{tikzpicture}[baseline={(a.south)}]
      \node (a) {$P$};
      \node (b) [right=0pt of a]{$M$};
      \node (c) [right=0pt of b]{$P$};
      \node (d) [right=0pt of c]{$M$};
      \node (e) [right=0pt of d]{$P$};
      \node (f) [right=0pt of e]{$M$};
      \path (a.north) edge [skip loop=8pt] (f.north);
      \path (b.north) edge [skip loop=8pt] (e.north);
      \path (c.north) edge [skip loop=4pt] (d.north);
    \end{tikzpicture} &
    \begin{tikzpicture}[baseline={(a.south)}]
      \node (a) {$P$};
      \node (b) [right=0pt of a]{$M$};
      \node (c) [right=0pt of b]{$P$};
      \node (d) [right=0pt of c]{$M$};
      \node (e) [right=0pt of d]{$P$};
      \node (f) [right=0pt of e]{$M$};
      \path (a.north) edge [skip loop=8pt] (f.north);
      \path (b.north) edge [skip loop=4pt] (e.north);
      \path (c.north) edge [skip loop=4pt] (d.north);
    \end{tikzpicture} &
    \begin{tikzpicture}[baseline={(a.south)}]
      \node (a) {$P$};
      \node (b) [right=0pt of a]{$M$};
      \node (c) [right=0pt of b]{$P$};
      \node (d) [right=0pt of c]{$M$};
      \node (e) [right=0pt of d]{$P$};
      \node (f) [right=0pt of e]{$M$};
      \path (a.north) edge [skip loop=8pt] (f.north);
      \path (b.north) edge [skip loop=4pt] (e.north);
      \path (c.north) edge [skip loop=8pt] (d.north);
    \end{tikzpicture} \\
    \begin{tikzpicture}[baseline={(a.south)}]
      \node (a) {$P$};
      \node (b) [right=0pt of a]{$M$};
      \node (c) [right=0pt of b]{$P$};
      \node (d) [right=0pt of c]{$M$};
      \node (e) [right=0pt of d]{$P$};
      \node (f) [right=0pt of e]{$M$};
      \path (a.north) edge [skip loop=8pt] (f.north);
      \path (b.north) edge [skip loop=4pt] (c.north);
      \path (d.north) edge [skip loop=8pt] (e.north);
    \end{tikzpicture} &
    \begin{tikzpicture}[baseline={(a.south)}]
      \node (a) {$P$};
      \node (b) [right=0pt of a]{$M$};
      \node (c) [right=0pt of b]{$P$};
      \node (d) [right=0pt of c]{$M$};
      \node (e) [right=0pt of d]{$P$};
      \node (f) [right=0pt of e]{$M$};
      \path (a.north) edge [skip loop=8pt] (f.north);
      \path (b.north) edge [skip loop=8pt] (c.north);
      \path (d.north) edge [skip loop=4pt] (e.north);
    \end{tikzpicture} &
    \begin{tikzpicture}[baseline={(a.south)}]
      \node (a) {$P$};
      \node (b) [right=0pt of a]{$M$};
      \node (c) [right=0pt of b]{$P$};
      \node (d) [right=0pt of c]{$M$};
      \node (e) [right=0pt of d]{$P$};
      \node (f) [right=0pt of e]{$M$};
      \path (a.north) edge [skip loop=4pt] (d.north);
      \path (b.north) edge [skip loop=8pt] (e.north);
      \path (c.north) edge [skip loop=8pt] (f.north);
    \end{tikzpicture} &
    \begin{tikzpicture}[baseline={(a.south)}]
      \node (a) {$P$};
      \node (b) [right=0pt of a]{$M$};
      \node (c) [right=0pt of b]{$P$};
      \node (d) [right=0pt of c]{$M$};
      \node (e) [right=0pt of d]{$P$};
      \node (f) [right=0pt of e]{$M$};
      \path (a.north) edge [skip loop=8pt] (d.north);
      \path (b.north) edge [skip loop=8pt] (e.north);
      \path (c.north) edge [skip loop=4pt] (f.north);
    \end{tikzpicture} \\
    \begin{tikzpicture}[baseline={(a.south)}]
      \node (a) {$P$};
      \node (b) [right=0pt of a]{$M$};
      \node (c) [right=0pt of b]{$P$};
      \node (d) [right=0pt of c]{$M$};
      \node (e) [right=0pt of d]{$P$};
      \node (f) [right=0pt of e]{$M$};
      \path (a.north) edge [skip loop=6pt] (d.north);
      \path (b.north) edge [skip loop=6pt] (e.north);
      \path (c.north) edge [skip loop=6pt] (f.north);
    \end{tikzpicture} & & &
  \end{tabular}
\end{center}
  \caption{Table of all contractions of the $\mathcal{O}\big(\kappa^6\big)$ term $\tr PMPMPM$}
  \label{tab:contractions}
\end{table}

At this stage we no longer need to distinguish between hops in positive or negative spatial 
directions. In contrast to
temporal hops, which get boosted by the baryon chemical potential, there is no asymmetry 
between them, and the gauge integration only depends on the number of links in a term.
We thus switch to a notation focussing on the dominant pairings,
\begin{align}
  &\tr \begin{tikzpicture}[
      baseline={([yshift=-2pt]one.center)},
      every node/.style = {anchor=south, text depth=0.35ex, inner sep=1pt}
    ]
    \node (one) {$X$};
    \node (two) [right=0pt of one] {$i$};
    \node (three) [right=0pt of two] {$Y$};
    \node (four) [right=0pt of three] {$i$};
  \end{tikzpicture} \;=\; 
\tr  \begin{tikzpicture}[
      baseline={([yshift=-2pt]one.center)},
      every node/.style = {anchor=south, text depth=0.35ex, inner sep=1pt}
    ]
    \node (one) {$X$};
    \node (two) [right=0pt of one] {$P$};
    \node (three) [right=0pt of two] {$Y$};
    \node (four) [right=0pt of three] {$M$};
    \path (two.north) edge [skip loop=6pt] (four.north);
    \node at ([yshift=6pt]$(two.north) !.5! (four.north)$) [scale=.75]{$i$};
  \end{tikzpicture} \;+\;
\tr  \begin{tikzpicture}[
      baseline={([yshift=-2pt]one.center)},
      every node/.style = {anchor=south, text depth=0.35ex, inner sep=1pt}
    ]
    \node (one) {$X$};
    \node (two) [right=0pt of one] {$M$};
    \node (three) [right=0pt of two] {$Y$};
    \node (four) [right=0pt of three] {$P$};
    \path (two.north) edge [skip loop=6pt] (four.north);
    \node at ([yshift=6pt]$(two.north) !.5! (four.north)$) [scale=.75]{$i$};
  \end{tikzpicture},
\end{align}
where $X$ and $Y$ symbolise the remainder of the term. Every contracted $P,M$ pair is labelled by an arbitrary number $i$, and the
terms are therefore invariant under relabelling.  The six pairings of $\tr PMPMPM$ in table~\ref{tab:contractions} are contained
in
\begin{equation} \label{eq:pmpmpm_pairing}
  \tr \tikz[baseline=-3pt] \matrix [matrix of math nodes,inner sep=1.5pt,ampersand replacement=\&] {1 \& 1 \& 2 \& 2 \& 3 \& 3 \\};, \;
  \tr \tikz[baseline=-3pt] \matrix [matrix of math nodes,inner sep=1.5pt,ampersand replacement=\&] {1 \& 2 \& 3 \& 3 \& 2 \& 1 \\};, \;
  \tr \tikz[baseline=-3pt] \matrix [matrix of math nodes,inner sep=1.5pt,ampersand replacement=\&] {1 \& 2 \& 3 \& 1 \& 2 \& 3 \\};.
\end{equation}
The number of equivalent labellings can be read from the notation, 
as the three terms have $2$, $3$ and $1$ distinct cyclic
permutation(s), respectively. In this notation the non-zero 
terms in \eq\eqref{eq:eff_action_expansion} read to leading order in $1/N_{\tau}$
\begin{align}
  \label{eq:combi}
  \exp \Bigg\{ -  \sum_{n=1}^{\infty} \frac{1}{n} \tr \big(P + M\big)^n \Bigg\} =&  \\
  1 - 
  \frac{1}{2}\tr \tikz[baseline=-3pt] \matrix [matrix of math nodes,inner sep=1.5pt,ampersand replacement=\&] {1 \& 1 \\}; + 
  \frac{1}{8}\tr \tikz[baseline=-3pt] \matrix [matrix of math nodes,inner sep=1.5pt,ampersand replacement=\&] {1 \& 1 \\};
  &\tr \tikz[baseline=-3pt] \matrix [matrix of math nodes,inner sep=1.5pt,ampersand replacement=\&] {2 \& 2 \\}; + 
  \frac{1}{4}\tr \tikz[baseline=-3pt] \matrix [matrix of math nodes,inner sep=1.5pt,ampersand replacement=\&] {1 \& 2 \\};
  \tr \tikz[baseline=-3pt] \matrix [matrix of math nodes,inner sep=1.5pt,ampersand replacement=\&] {1 \& 2 \\}; -
  \frac{1}{2}\tr \tikz[baseline=-3pt] \matrix [matrix of math nodes,inner sep=1.5pt,ampersand replacement=\&] {1 \& 1 \& 2 \& 2 \\};
  + \mathcal{O}\big(\kappa^6,\frac{1}{N_{\tau}}\big)\;. \nn
\end{align}
The combinatorial prefactors $1/g$ of the trace products are determined by 
the symmetries of the individual terms. We have
\begin{equation}
 \frac{1}{g}=    \frac{%
    \tikz \node [scale=0.75,align=center] {\# of unique cyclic \\ permutations of the traces};
  }{%
 \tikz \node {$n_2!n_4!\cdots{}n_N!\, 2^{n_2} 4^{n_4} \cdots N^{n_N}$};} \;.
\end{equation}
The numerator is the number of cyclic permutations within all traces that stay 
different under relabelling. 
The $n_i$ in the denominator is the
number of trace factors over $i$ matrices (e.g.~the third term in (\ref{eq:combi}) has 
$n_2=2$ and the fifth $n_4=1$),  and $N$ is the total number of 
matrix factors (or the order of $\kappa$) of the term. 

\subsection{The terms contributing to the chain}

To start the chain
we define an \emph{open end} to consist of two consecutive hops that are paired, such as
``$1\,1$''. In \eq\eqref{eq:pmpmpm_pairing} we see that the first term has three such open ends, 
the second has two and the final has
no open ends. These open ends turn  into $W_{1,1}$ terms in the final expressions and are
therefore the attachment points for building a chain.
A new element of the chain is added by inserting a new open end between the pairing. 
Instead of doing a hop forward and backward in the pair, this corresponds to taking a detour
through the new point
\begin{subequations}
  \begin{align}
    \cdots \tikz[baseline=-3pt] \matrix [matrix of math nodes,inner sep=1.5pt,ampersand replacement=\&] {1 \& 1 \\}; \hspace{.25cm}
    &\tikz \draw[->,>=stealth] (0,0) -- +(1cm,0); \hspace{.3cm} \cdots \: W_{1,1}(\vec{x}), \\
    \cdots \tikz[baseline=-3pt] \matrix [matrix of math nodes,inner sep=1.5pt,ampersand replacement=\&] {1 \& 2 \& 2 \& 1 \\}; \hspace{.25cm}
    &\tikz \draw[->,>=stealth] (0,0) -- +(1cm,0); \hspace{.3cm} \cdots \: W_{2,1}(\vec{x}) W_{1,1}(\vec{x} + i), \\
    \cdots \tikz[baseline=-3pt] \matrix [matrix of math nodes,inner sep=1.5pt,ampersand replacement=\&] {1 \& 2 \& 3 \& 3\& 2 \& 1 \\}; \hspace{.25cm}
    &\tikz \draw[->,>=stealth] (0,0) -- +(1cm,0); \hspace{.3cm} \cdots \: W_{2,1}(\vec{x}) W_{2,1}(\vec{x} + i) W_{1,1}(\vec{x} + i + j).
  \end{align}
\end{subequations}
The prefactors of terms in this resummation can be calculated from symmetry arguments. 
Assume that we know the symmetry prefactor $1/g$ of
a term with $N$ open ends. 
Extending one of these can be done in $N/g$ distinct ways, 
which all break the previous symmetry. The sum of all
such insertions thus have a prefactor of $N/g$ which is $N$ times that of the base diagram. 
Instead of counting the number of
permutations we can think of the number of ways to add $n$ links to $N$ open ends. 
The total combinatorial factor for a graph with $N$ open ends, an internal
symmetry of $g$ and with $n$ link insertions is therefore
\begin{equation} \label{eq:resummed_symmetry}
  \frac{1}{g} \binom{N - 1 + n}{n}\;.
\end{equation}

\subsection{Dirac traces}

Next we will compute the trace over the spin indices and show that they give a simple contribution to leading order in $1/N_{\tau}$.
Comparing the expression for the static propagator \eqref{eq:static_propagator} with the definition of a contraction
\eqref{eq:contraction_def} one sees that the contractions do not constrict $\tau_1$ and $\tau_2$. Therefore when summing independently
over these, terms with $\tau_1=\tau_2$ are subleading in $N_{\tau}$ and we can drop the $\delta_{\tau_1,\tau_2}$ term,
\begin{equation} \label{eq:static_propagator_simple}
  T_{ab}^{\alpha\beta}(x,y) 
  \begin{tikzpicture}[baseline={([yshift=-1.5pt]equal.center)}]
    \node (equal) {$=$};
    \node[above=0pt of equal] [scale=0.65,align=center] {leading\\[0pt]order};
  \end{tikzpicture}
  \big(1 - \gamma_0\big)^{\alpha\beta} B_{ab}(x,y)\;.
\end{equation}
The only exception to this is a contraction
\begin{align}
  &\begin{tikzpicture}[
      baseline = .5ex,
      every node/.style = {anchor=south, text depth=0.35ex, inner sep=1pt}
    ]
    \node (one) {$P(x,y)$};
    \node (two) [right=0pt of one]{$M(y,z)$};
    \path ([xshift=7pt]one.north west) edge [skip loop=6pt] ([xshift=7pt]two.north west);
  \end{tikzpicture}
  =\begin{tikzpicture}[
      baseline = .5ex,
      every node/.style = {anchor=south, text depth=0.35ex, inner sep=1pt}
    ]
    \node (one) {$P(x,y)$};
    \node (two) [right=0pt of one]{$M(y,z)$};
    \node (three) [right=0pt of two]{$\delta_{\tau_y,\tau_z}$};
  \end{tikzpicture}
\end{align}
but the $\delta_{\tau_1,\tau_2}$ in \eqref{eq:static_propagator} would lead to backtracking in the $i$ direction, which vanishes
after the gamma traces.  Hence the Dirac structure of any trace term in the cold region is always proportional to $B$ and has the
general structure
\begin{equation}
  \tr \left[ (1-\gamma_0)(1 \pm \gamma_i) (1-\gamma_0) (1 \pm \gamma_j) \cdots \right]\;,
\end{equation}
where every $P$ contribute with a $(1-\gamma_0)(1-\gamma_i)$ pair and every $M$ with a $(1-\gamma_0)(1+\gamma_i)$ pair.  To
shorten the calculation, we introduce an intermediate notation $g_{\mu} = (1 - \gamma_{\mu})$ and $\bar{g}_{\mu} = (1 +
\gamma_{\mu})$. Picking the $P$ contribution for now, the above expression reads
\begin{equation}
  \tr \left[ g_0 g_i g_0 \cdots \right].
\end{equation}
Expanding the first two terms, we get
\begin{align}
  \tr\left[(1 - \right.&\left.\gamma_0 - \gamma_i + \gamma_0 \gamma_i)g_0  \cdots\right] \nonumber\\
  &= \tr\left[g_0 \cdots\right] - \tr\left[\gamma_0g_0 \cdots\right]
  - \tr\left[\gamma_i g_0 \cdots\right] + \tr\left[\gamma_0\gamma_ig_0 \cdots\right]
\end{align}
Using the Dirac identities we know that $\gamma_0$ and $\gamma_i$ anti-commute, and we can easily see that
$\gamma_0 g_0 = \gamma_0 (1-\gamma_0) = (\gamma_0 - 1) = -g_0$. Inserting this into the above expression results in
\begin{align}
  \tr\left[g_0 \cdots\right] + \tr\left[g_0 \cdots\right]
  - \tr\left[\gamma_i g_0 \cdots\right] + \tr\left[\gamma_ig_0 \cdots\right]
  = 2 \tr\left[ g_0  \cdots \right].
\end{align}
The same calculation holds for $g_0 \bar{g}_i$ pairs, and we can thus replace every $g_0 g_i$ pair with a factor 2 until there is
only one pair left, the trace of which is 4, or the dimension of the system
\begin{equation}
  \tr \big[ \underbrace{g_0 g_i g_0 \cdots}_{n \text{ pairs}} \big] = 2^{n-1} \tr \big[g_0 g_i \big] = 2^{n+1}.
\end{equation}

\subsection{Recursive gauge integration for the chain}

Finally we will compute the spatial link integrals for the chain and see that we reproduce the substitution in
\eq\eqref{eq:chain_subst}. We argued that chain of length $n$ can be represented as 
\begin{equation}
  \tikz[baseline=-3pt] \matrix [matrix of math nodes,inner sep=1.5pt,ampersand replacement=\&] {1 \& 2 \& 3 \& 4 \& \dots \& n \& n \& \dots \& 4 \& 3 \& 2 \& 1 \\};,
\end{equation}
an assumption we will finally settle in this section.  The expression for the chain has a recursive structure, and it is therefore
natural to define the matrices $G_m$ such that
\begin{center}
  \begin{tikzpicture}
    \matrix (nodes) [matrix of math nodes,inner sep=1.5pt,ampersand replacement=\&] {1 \& 2 \& 3 \& 4 \& \dots \& n \& n \& \dots \& 4 \& 3 \& 2 \& 1 \\};
    \draw [decorate,decoration={brace,mirror,amplitude=5pt}]
      ([yshift=-.25]nodes-1-1.south) -- ([yshift=-.25]nodes-1-12.south)
      node[midway,below=5pt,scale=.75] {$G_n$};
    \draw [decorate,decoration={brace,mirror,amplitude=5pt}]
      ([yshift=-.85cm]nodes-1-2.south) -- ([yshift=-.85cm]nodes-1-11.south)
      node[midway,below=5pt,scale=.75] {$G_{n-1}$};
    \draw [decorate,decoration={brace,mirror,amplitude=2pt,raise=4pt}]
      ([yshift=-1.45cm]nodes-1-6.south) -- ([yshift=-1.45cm]nodes-1-7.south)
      node[midway,below=5pt,scale=.75] {$G_{1}$};
  \end{tikzpicture}
\end{center}
$G_n$ is then defined in terms of $G_{n-1}$ such that
\begin{align}
  G_{n}^{af}(\tau_1,\tau_2;\vec{x}_0) &= 2 \kappa^2 \sum_{i_0,\tau_3} \int \mathrm{d} U_{\vec{x}_0,i_0}(\tau_2) \;
    B_{\vec{x}_0}^{ab}(\tau_1,\tau_2)U_{\vec{x}_0,i_0}^{bc}(\tau_2) \nonumber \\
    &\hspace{3cm}\times G_{n-1}^{cd}(\tau_2,\tau_3;\vec{x}_0+i_0) B^{de}_{\vec{x}_0+i_0}(\tau_3,\tau_2)
    U_{\vec{x}_0,i_0}^{\dagger,ef}(\tau_2) \nonumber\\
  &= \frac{2 \kappa^2}{N_c} \sum_{i_0,\tau_3} B_{\vec{x}_0}^{ab}(\tau_1,\tau_2) G_{n-1}^{cd}(\tau_2,\tau_3;\vec{x}_0+i)
    B^{de}_{\vec{x}_0+i}(\tau_3,\tau_2) \delta_{ce}\delta_{bf} \nonumber\\
  &= \frac{2 \kappa^2}{N_c} \sum_{i_0,\tau_3} B_{\vec{x}_0}^{af}(\tau_1,\tau_2) \tr_c \Big[ G_{n-1}(\tau_2,\tau_3;\vec{x}_0+i)
    B_{\vec{x}_0+i}(\tau_3,\tau_2) \Big] \;.\label{eq:chain_integral}
\end{align}
Here $\vec{x}_0$ is the coordinate of the starting pair of the chain and we
see a recursive structure for the spatial positions $\vec{x}_{m+1} = \vec{x}_m + i_m$ of the chain's
end. $G_{n-1}$ is of course in
turn defined in terms of $G_{n-2}$,
\begin{equation}
  G_{n-1}^{ab}(\tau_2,\tau_3;\vec{x}_1) = \frac{2 \kappa^2}{N_c} \sum_{i_1,\tau_4} B_{\vec{x}_1}^{ab}(\tau_2,\tau_3)
    \tr_c \Big[ G_{n-2}(\tau_3,\tau_4;\vec{x}_1+i_1) B_{\vec{x}_1+i_1}(\tau_4,\tau_3) \Big]\;.
\end{equation}
Inserting the expression for $G_{n-1}$ into the expression for $G_{n}$ we get
\begin{align}
  G_{n}^{af}(\tau_1,\tau_2;\vec{x}_0) = \bigg(\frac{2 \kappa^2}{N_c}\bigg)^2 \sum_{\tau_3,\tau_4} \sum_{i_0,i_1} B_{\vec{x}_0}^{af}&(\tau_1,\tau_2)
    \tr_c \Big[ G_{n-2}(\tau_3,\tau_4;\vec{x}_1+i_1) B_{\vec{x}_1+i_1}(\tau_4,\tau_3) \Big] \nonumber \\
  &\times\underbrace{tr_c \Big[ B_{\vec{x}_0+i_0}(\tau_2,\tau_3)
    B_{\vec{x}_0+i_0}(\tau_3,\tau_2)\Big]}_{-\frac{1}{2}W_{2,1}(\vec{x}_0+i_0)}\;,
\end{align}
which has the exact same structure as \eq\eqref{eq:chain_integral} except that we have a factor of $W_{2,1}$ and $G_{n-1}$ has
been replaced by $G_{n-2}$. The recursion ends when we are at $G_1$, which is the open end and has the slightly different form
\begin{align}
  G_{1}^{ae}(\tau_1,\tau_2;\vec{x}_n) &= 2 \kappa^2 \sum_{i_n} \int \mathrm{d} \vec{U}_{\vec{x}_n,i_n} \, B^{ab}_{\vec{x}_n}(\tau_1,\tau_2)
  U_{\vec{x}_n,i_n}^{bc}(\tau_2) B^{cd}_{\vec{x}_n+i_n}(\tau_2,\tau_2) U_{\vec{x}_n,i_n}^{\dagger,de}(\tau_2) \nonumber \\
  &= \frac{2 \kappa^2}{N_c} \sum_{i_n}B^{ae}_{\vec{x}_n}(\tau_1,\tau_2)
  \underbrace{\tr_c \Big[ B_{\vec{x}_n+i_n}(\tau_2,\tau_2) \Big]}_{\frac{1}{2} W_{1,1}(\vec{x}_n + i_n)}\;. \label{eq:open_link_integral}
\end{align}
The final result for $G_n$ is therefore
\begin{align}
  G_{n}^{ab}(\tau_1,\tau_2;\vec{x}_0) &= B_{\vec{x}_0}^{ab}(\tau_1,\tau_2) \bigg(\frac{2 \kappa^2}{N_c}\bigg)^n \sum_{\substack{\tau_3,\tau_4,\\\dots,\tau_{n+1}}}
  \sum_{\substack{i_0,i_1,\\\dots,i_n}}W_{1,1}(\vec{x}_n + i_n) \prod_{k=2}^{n} \big(-W_{2,1}(\vec{x}_k)\big) \nonumber \\
  &=B_{\vec{x}_0}^{ab}(\tau_1,\tau_2) \bigg(\frac{2 \kappa^2}{N_c}\bigg)^n N_{\tau}^{n-1}
  \sum_{\substack{i_0,i_1,\\\dots,i_n}}W_{1,1}(\vec{x}_n + i_n) \prod_{k=2}^{n} \big(-W_{2,1}(\vec{x}_k)\big)\;, \label{eq:full_gn}
\end{align}
where the sum over the temporal variables could be trivially evaluated as the $W_{n,m}$'s are independent of their time argument.

To tie it all together let us consider a generic contribution which has $N$ open ends,
\begin{equation} \label{eq:chain_begin}
  \mathcal{C}_0 = \frac{1}{g} \tr \big[ G_1(\tau_1,\tau_2;\vec{x}_1) M_1 G_1 (\tau_3,\tau_4;\vec{x}_2) M_2 \cdots G_1 (\tau_{2N-1},\tau_{2N};\vec{x}_N) M_N  \big]\;,
\end{equation}
where the matrices $M_i$ are the rest of the term, comparable to the left hand side of \eq\eqref{eq:chain_schematic}. Inserting
the expression for $G_1$ gives
\begin{align}
  \mathcal{C}_0 = \frac{1}{g} \tr \big[ &B(\tau_1,\tau_2;\vec{x}_1) M_1 B(\tau_3,\tau_4;\vec{x}_2) M_2 \cdots B(\tau_{2N-1},\tau_{2N};\vec{x}_N) M_N  \big]
  \nonumber\\
  &\times \bigg(\frac{\kappa^2}{N_c}\bigg)^N\;\sum_{\mathclap{i_1,i_2,\dots,i_N}} W_{1,1}(\vec{x}_1 + i_1) W_{1,1}(\vec{x}_2 + i_2) 
    \cdots W_{1,1}(\vec{x}_N + i_N).
\end{align}
We can now attach chains of length $n_i$ to each of the $N$ open ends so that the total length of the chain is $n$, corresponding
to the right hand side of \eq\eqref{eq:chain_schematic}
\begin{align}
  \mathcal{C}_n = \sum_{\mathclap{n_1,n_2,\dots,n_N}} \;\quad \frac{1}{g_{\{n_i\}}} \tr \big[ G_{n_1 + 1}(\tau_1,\tau_2;\vec{x}_1) M_1
  &G_{n_2+1} (\tau_3,\tau_4;\vec{x}_2) M_2 \nonumber\\ 
  \cdots G_{n_N + 1} (\tau_{2N-1},&\tau_{2N};\vec{x}_N) M_N  \big] \delta\bigg(\sum_{i=1}^N n_i - n\bigg)\;.
\end{align}
This gives a symmetry factor that depends on the partitioning of the attachments $\{n_i\}$.
We insert the expression for $G_n$ from \eq\eqref{eq:full_gn}, which gives us
\begin{align}
  \mathcal{C}_n =& \sum_{\mathclap{n_1,n_2,\dots,n_N}} \;\quad \frac{1}{g_{\{n_i\}}} \tr \big[ B(\tau_1,\tau_2;\vec{x}_1) M_1 B(\tau_3,\tau_4;\vec{x}_2) M_2 \cdots
  B(\tau_{2N-1},\tau_{2N};\vec{x}_N) M_N  \big] \nonumber\\
  &\times \bigg(\frac{\kappa^2}{N_c}\bigg)^{N+\mathrlap{n}} N_{\tau}^n \sum_{\mathrm{dof}} \prod_{j=1}^N W_{1,1}(\vec{x}_{j n_j} + i_{j n_j})
    \prod_{k=0}^{n_j} \big(-W_{2,1}(\vec{x}_{j k})\big) \;\delta\bigg(\sum_{i=1}^N n_i - n\bigg)\;,
\end{align}
where $\vec{x}_{j k}$ is the $k$'th position of the chain originating from the $j$'th open end, corresponding to $\vec{x}_j$ from
\eq\eqref{eq:chain_begin}. The degrees of freedom are the directions the hops can take.  We see that the base diagram, what is
explicitly left in the trace, is the same for the term with and without attachments. There is the integral over temporal gauge
links which we evaluate by embedding this term onto a skeleton cluster expansion graph with the same geometry.  We sum over the
subclass of non-overlapping chains, meaning that the coordinates $\vec{x}_{j k}$ never overlap with each other, nor the positions
of the base diagram. This subclass of terms is labelled $\mathcal{C}^{\ast}$.  The integrals over the remaining temporal gauge
links therefore factorise and we can sum all partitions of the attachments $\{n_i\}$ into one term. The sum of the symmetry
factors of the partitions is exactly what we computed in \eq\eqref{eq:resummed_symmetry},
\begin{equation}
  \sum_{\mathclap{n_1,n_2,\dots,n_N}} \;\quad \frac{1}{g_{\{n_i\}}} = \frac{1}{g} \binom{N-1+n}{n}\;,
\end{equation}
which means that the integral over this embedding, $\mathcal{C}^{\ast}$, is
\begin{align}
  \int \mathrm{D} W \, &\mathcal{C}_n^{\ast} = \frac{1}{g} \binom{N-1+n}{n} \Omega \int \mathcal{D} W \, \det\big[Q_{\mathrm{stat}}\big]
    \tr \big[ B_1 M_1 B_2 M_2 \cdots B_N M_N  \big] \nonumber\\
  &\times \bigg(\frac{2d \kappa^2}{N_c}\int \mathrm{d} W \, \det\big[Q_{\mathrm{stat}}\big] W_{1,1}\bigg)^N
    \bigg(-\frac{2d \kappa^2N_{\tau}}{N_c}\int \mathrm{d} W \, \det\big[Q_{\mathrm{stat}}\big] W_{2,1}\bigg)^{\mathrlap{n}}\;.
\end{align}
Here $\Omega$ is the embedding factor of the base diagram and every factor in the chain brings a lattice embedding of $2d$ as argued
in section \ref{sec:linked_cluster}. Finally we can sum over the total length of the attachments, $n$, and for brevity we only include the
$n$-dependent factors of the previous expression,
\begin{align}
  \sum_{n=0}^{\infty} \binom{N-1+n}{n} \bigg(\underbrace{-\frac{2d \kappa^2N_{\tau}}{N_c} \vrule width0pt height0pt depth3ex\relax
    \int \mathrm{d} W \, \det\big[Q_{\mathrm{stat}}\big] W_{2,1}}_{-(2d) h_2 I_{2,1}\strut}\bigg)^{\mathrlap{n}}& \nonumber \\
  &\hspace*{-2cm}= \bigg(\frac{1}{1 + (2d) h_2 I_{2,1}}\bigg)^N\;,
\end{align}
which is the same as the resummation formula proposed in \eq\eqref{eq:chain_resummation_formula}.

\end{appendices}

\end{document}